\documentclass[USenglish,oneside,twocolumn]{article}

\usepackage[utf8]{inputenc}
\usepackage[big]{dgruyter_NEW}

\usepackage[table,xcdraw]{xcolor}

\definecolor{dark}{gray}{0.90}
\definecolor{white}{RGB}{169,208,142}
\definecolor{medium}{RGB}{169,208,142}

\usepackage{array}
\usepackage{subfigure}
\usepackage{yfonts,xspace}
\usepackage{ragged2e}
\usepackage{algorithm}
\usepackage{algorithmic}
\usepackage{appendix}
\usepackage{comment}
\usepackage{hyperref}

\usepackage{booktabs}
\usepackage{multirow}
\usepackage{longtable}

\usepackage{upquote}
\usepackage{listings}

\usepackage{soul, color}
\definecolor{lightgray}{rgb}{0.95, 0.95, 0.95}
\definecolor{darkgray}{rgb}{0.4, 0.4, 0.4}
\definecolor{editorGray}{rgb}{0.95, 0.95, 0.95}
\definecolor{editorOcher}{rgb}{1, 0.5, 0} 
\definecolor{editorGreen}{rgb}{0, 0.5, 0} 
\definecolor{orange}{rgb}{1,0.45,0.13}		
\definecolor{olive}{rgb}{0.17,0.59,0.20}
\definecolor{brown}{rgb}{0.69,0.31,0.31}
\definecolor{purple}{rgb}{0.38,0.18,0.81}
\definecolor{lightblue}{rgb}{0.1,0.57,0.7}
\definecolor{lightred}{rgb}{1,0.4,0.5}
\lstdefinelanguage{JavaScript}{
  morekeywords={typeof, window, new, true, false, catch, function, return, null, catch, switch, var, if, in, else, case, break, document, write, createElement, width, height, display, visibility, border, document.write},
  morecomment=[s]{/*}{*/},
  morecomment=[l]//,
  morestring=[b]",
  morestring=[b]'
}

\lstdefinelanguage{HTML5}{
  language=html,
  sensitive=true,	
  alsoletter={<>=-},	
  morecomment=[s]{<!-}{-->},
  tag=[s],
  otherkeywords={
  >,
	<!DOCTYPE,
  </html, <html, <head, <title, </title, <style, </style, <link, </head, <meta, />,
	</body, <body,
	</div, <div, </div>,
	</p, <p, </p>, <a, <h1, </h1>, </h1,
	</script, <script,
  <canvas, /canvas>, <svg, <rect, <animateTransform, </rect>, </svg>, <video, <source, <iframe, </iframe>, </video>, <image, </image>, <header, </header, <article, </article, </iframe, <img, <span, </span, </a,<button, </button>, <ul>, </ul>, <li>, </li>
  },
  ndkeywords={
  =, ===, ==,
  charset=, src=, id=, width=, height=, style=, type=, rel=, href=, name=, tabindex=, align=, scrolling=, framespacing=, frameborder=, allowtransparency=, data-dm-title=, data-dm-format=, data-dm-filesize=, target=, data-dm=, data-dm-icon=, data-dm-href-free=, data-dm-filename=, data-dm-hosted-file=, data-dm-href= , data-dm-carregado=, class=, alt=,
  fill=, attributeName=, begin=, dur=, from=, to=, poster=, controls=, x=, y=, repeatCount=, xlink:href=,
  margin:, padding:, background-image:, border:, top:, left:, position:, width:, height:, margin-top:, margin-bottom:, font-size:, line-height:,
  transform:, -moz-transform:, -webkit-transform:,
  animation:, -webkit-animation:,
  transition:,  transition-duration:, transition-property:, transition-timing-function:,
  }
}

\lstdefinestyle{htmlcssjs} {%
  backgroundcolor=\color{editorGray},
  basicstyle=\fontsize{8}{8}\ttfamily,
  frame=tb,
  captionpos=b,
  belowcaptionskip=\medskipamount,
  xleftmargin={0.5cm},
  numbers=left,
  stepnumber=1,
  firstnumber=1,
  numberfirstline=true,	
  identifierstyle=\color{black},
  keywordstyle=\color{blue}\ttfamily,
  ndkeywordstyle=\color{editorGreen}\ttfamily,
  stringstyle=\color{black}\ttfamily,
  commentstyle=\color{brown}\ttfamily,
  language=HTML5,
  alsolanguage=JavaScript,
  alsodigit={.:;},	
  tabsize=2,
  showtabs=false,
  showspaces=false,
  showstringspaces=false,
  extendedchars=true,
  breaklines=true,
  numberstyle=\tiny\color{gray},
  literate=%
  {Ö}{{\"O}}1
  {Ä}{{\"A}}1
  {Ü}{{\"U}}1
  {ß}{{\ss}}1
  {ü}{{\"u}}1
  {ä}{{\"a}}1
  {ö}{{\"o}}1
}

	\journalname{Proceedings on Privacy Enhancing Technologies}
	\DOI{Editor to enter DOI}
	\startpage{1}
 	\received{..}
	\revised{..}
	\accepted{..}

	\journalyear{..}
	\journalvolume{..}
	\journalissue{..}

\newcommand{\name}{\textsc{FP-Radar}\xspace}
\newcommand{\WatchDog}{\textsc{FP-Radar}\xspace} 
\newcommand\blfootnote[1]{%
  \begingroup
  \renewcommand\thefootnote{}\footnote{#1}%
  \addtocounter{footnote}{-1}%
  \endgroup
}

\begin{document}

\title{\huge \WatchDog: Longitudinal Measurement and Early Detection of Browser Fingerprinting}

\runningtitle{\WatchDog: Longitudinal Measurement and Early Detection of Browser Fingerprinting}


\author[]{Pouneh Nikkhah Bahrami*} \blfootnote{\textbf{*Corresponding Author: Pouneh Nikkhah Bahrami}, University of California, Davis, Email: pnikkhah@ucdavis.edu}
\author[]{Umar Iqbal} \blfootnote{\textbf{Umar Iqbal}, University of Washington, Email:  umar@cs.washington.edu}
\author[]{Zubair Shafiq} \blfootnote{\textbf{Zubair Shafiq}, University of California, Davis, Email: zubair@ucdavis.edu}

\begin{abstract}
{
Browser fingerprinting is a stateless tracking technique that attempts to combine information exposed by multiple different web APIs to create a unique identifier for tracking users across the web. 
Over the last decade, trackers have abused several existing and newly proposed web APIs to further enhance the browser fingerprint. 
Existing approaches are limited to detecting a specific fingerprinting technique(s) at a particular point in time. 
Thus, they are unable to systematically detect novel fingerprinting techniques that abuse different web APIs. 
In this paper we propose \name, a machine learning approach that leverages longitudinal measurements of web API usage on top-100K websites over the last decade, for early detection of new and evolving browser fingerprinting techniques. 
The results show that \name is able to early detect the abuse of newly introduced properties of already known (e.g., \texttt{WebGL}, \texttt{Sensor}) and as well as {previously unknown} (e.g., \texttt{Gamepad}, \texttt{Clipboard}) APIs for browser fingerprinting. 
To the best of our knowledge, \name is also first to detect the abuse of the \texttt{Visibility} API for ephemeral fingerprinting in the wild.
}
\end{abstract}
\maketitle
\vspace{-.6in}
\section{Introduction}
\label{sec: introduction}

The online tracking ecosystem employs increasingly sophisticated tracking techniques to track users across the web \cite{Lerner16InternetJonesUSENIX,Englehardt16MillionSiteMeasurementCCS,Merzdovnik17BlockMeIfYouCanESP,Yu16TrackingTheTrackersWWW}. 
In addition to well-known \textit{stateful tracking} using third-party cookies, trackers have now started to use more intrusive \textit{stateless tracking} techniques such as browser fingerprinting to gather device-specific identifying information captured through various HTTP header fields and APIs \cite{Fifield2015FCfontmetrics,olejnik2015leaking,mowery2011fingerprinting,mowery2012pixel,cao2017cross,Englehardt16MillionSiteMeasurementCCS,laperdrix2020browserfpsurvey}.
Stateless tracking is more intrusive than stateful tracking because the former does not lend itself to transparency and control. 
While cookies are directly observed and removed at the client-side, browser fingerprint is not directly visible at the client-side and it cannot be trivially removed or even modified.
As web browsers have started to implement aggressive countermeasures against stateful tracking \cite{safarifullthirdpartycookieblockingJohnWder,wood2019firefox69release,schuh2020cookiesobselete}, it has encouraged trackers to migrate to more opaque and invasive stateless tracking \cite{schuh2020cookiesobselete,iqbal21fpinspector}.

Browser fingerprinting techniques have evolved over time. 
As web browsers support new functionality by adding new APIs or update existing APIs \cite{snyder2016browser}, the browser's fingerprinting surface has continued to expand.
Early work by Mayer \cite{mayer2009pamphleteer} and Eckersley \cite{eckersley2010unique} demonstrated simple fingerprinting techniques that abuse information exposed in HTTP headers and a few APIs. 
A steady stream of more sophisticated fingerprinting techniques have since been developed that abuse existing and new APIs. 
For example, researchers have shown that \texttt{Canvas} \cite{mowery2012pixel}, \texttt{WebGL} \cite{cao2017cross}, fonts \cite{Fifield2015FCfontmetrics}, extensions \cite{Starov2017XHound}, the \texttt{Audio} API \cite{Englehardt16MillionSiteMeasurementCCS}, the \texttt{Battery Status} API \cite{olejnik2015leaking,Olejnik17BatteryStatusIWPE}, the \texttt{Performance} API \cite{sanchez2018clock}, and even sensor APIs \cite{bojinov2014mobile,Das18MobileSensorsCCS} can expose information that can be abused to build a more reliable fingerprint.
Thus, as new APIs are introduced in web browsers, it is reasonable to expect that they might be abused to implement novel browser fingerprinting techniques.
In summary, browser fingerprinting is not a static phenomenon, but it is rather evolving; as novel fingerprinting techniques are designed over time.

Browser fingerprinting and its privacy implications have received much attention from the research community.
Researchers have conducted large-scale measurements to study the prevalence of browser fingerprinting  \cite{nikiforakis2013cookieless,acar2013fpdetective,Acar14WebNeverForgetsCCS,faizkhademi2015fpguard,Englehardt16MillionSiteMeasurementCCS,Olejnik17BatteryStatusIWPE,Das18MobileSensorsCCS,fowlerWashingtonPostFingerprinting,iqbal21fpinspector}. 
However, prior research on browser fingerprinting is lacking in two major ways.
First, prior work is mostly limited to analyzing a specific fingerprinting technique(s) at a particular point in time.
Since fingerprinting techniques evolve over time, it is important to study browser fingerprinting longitudinally. 
Second, prior work is limited to detecting deployment of already known fingerprinting techniques.
It is important to detect new fingerprinting techniques in a timely fashion because early detection can aid proactive mitigation efforts by the standards bodies \cite{w3cFPGuidance} and also prompt deployment of targeted countermeasures by browser vendors \cite{TOR_FIngerprintProtection,privacyBudget}.

We propose \WatchDog, a machine learning approach for early detection of web API abuse for fingerprinting. 
\WatchDog detects abuse of new methods of existing APIs or new APIs altogether by using the guilt-by-association principle. 
More specifically, it first uses the Wayback Machine to crawl the historical snapshots of  scripts on top-100K websites over the last decade. 
\WatchDog conducts static analysis to construct a series of temporal API co-occurrence graphs for each year.
\WatchDog then uses hand-crafted and embedding features to predict the evolution of co-occurrence relationships between different API keywords over the years. 
%
\WatchDog then builds and labels temporal clusters, including the fingerprinting cluster, using the temporal graphs.
Finally, \WatchDog tracks the membership of the fingerprinting cluster over time for early detection of API abuse.

The results show that \WatchDog is able to detect the abuse of already known as well as unknown APIs for fingerprinting. 
First, \WatchDog detects the abuse of a number of previously unknown APIs including \texttt{Page Visibility}, \texttt{Gamepad}, \texttt{Clipboard}, and \texttt{Network Information} for browser fingerprinting. 
We find novel types of user environment/hardware fingerprinting such as peripheral configuration via \texttt{Gamepad} and system capabilities via \texttt{Network Information} APIs. 
We also find that even though an API (e.g., \texttt{Page Visibility}) does not directly expose highly identifying information it can be abused for ephemeral fingerprinting. 
To the best of our knowledge, \name is also the first to detect the abuse of web APIs for ephemeral fingerprinting in the wild. 
Second, \WatchDog detects the abuse of newly introduced features of APIs that are already known to be abused for fingerprinting. 
We find that several of the newly introduced features of \texttt{Navigator} (e.g., related to hardware capabilities such as memory), \texttt{Performance} (e.g., time for DNS lookup and page rendering), and \texttt{WebGL} (e.g., WebGL2 capabilities) are now being abused for fingerprinting. 
Finally, \WatchDog is able to detect the fingerprinting abuse of APIs before/at their disclosure or at the time of their release by browser vendors or their first occurrence in our data. 
We find that \WatchDog's time-to-detection is often several years before public disclosure (e.g., as much as 6 years for \texttt{Gamepad} and 7 years for \texttt{Page Visibility} APIs).

We summarize our key contributions as follows: 

\begin{enumerate}
    \item A retrospective \textbf{longitudinal measurement} study of web API usage over the last decade. 

    \item A graph-based \textbf{supervised ML approach} that builds a series of API co-occurrence graphs to predict the evolution of API usage in the future. 
    
    \item A graph-based
    \textbf{unsupervised ML approach} that clusters temporal API co-occurrence graphs for early detection of their abuse for fingerprinting. 
\end{enumerate}



\section{Background \& Related Work}
\label{sec: related work}

\subsection{Background}
Web browsers support standardized web APIs to facilitate  feature-rich websites that can be seamlessly loaded on different browsers (e.g., Chrome, Firefox), operating systems (e.g., Mac/Windows), and devices (e.g., mobile/desktop). 
Unfortunately, the rich set of information exposed by the web APIs can also be exploited by trackers to fingerprint users' devices. 
Trackers can simply combine several pieces of readily available information, such as the operating system name, browser name, browser version in the user-agent field, to build a fingerprint that can distinguish between different devices.
Trackers can also use more sophisticated fingerprinting techniques that exploit subtle differences in the underlying hardware/software configurations and capabilities to gather distinctive information. 
For example, canvas images are rendered differently on different browsers due to the differences in their hardware/software image processing pipeline. 
Combining several of these fingerprinting techniques, trackers can create a fingerprint that is often sufficient to uniquely and persistently identify the web browser \cite{eckersley2010unique}.

Browser fingerprinting is called stateless tracking since there is no need to store state at the client-side, as done in traditional cookie-based stateful tracking. 
Stateless tracking is considered more intrusive than stateful tracking because the former does not lend itself to transparency and control. 
While cookies and other types of client-side storage mechanisms (e.g., localStorage, IndexedDB) can be readily observed and blocked at the client-side, browser fingerprint is not directly visible at the client-side and it cannot be trivially removed or even modified.
As web browsers have started to implement aggressive countermeasures against stateful tracking \cite{safarifullthirdpartycookieblockingJohnWder,wood2019firefox69release,schuh2020cookiesobselete}, it has encouraged trackers to migrate to more opaque and invasive stateless tracking \cite{schuh2020cookiesobselete,iqbal21fpinspector}.
Browser fingerprinting is already being used for cross-site tracking \cite{iqbal21fpinspector, Alaca16ACSACDeviceFingerprinting, Laperdrix19DIMVAMorellian} and is universally regarded as an abusive practice by standards bodies \cite{w3cFPGuidance, tagtracking} and web browsers \cite{appleFingerprintingBlogPost, privacyBudget, mozillaFingerprintingBlogPost}.

\begin{figure*}[ht]
  \centering
  \includegraphics[width=\textwidth]{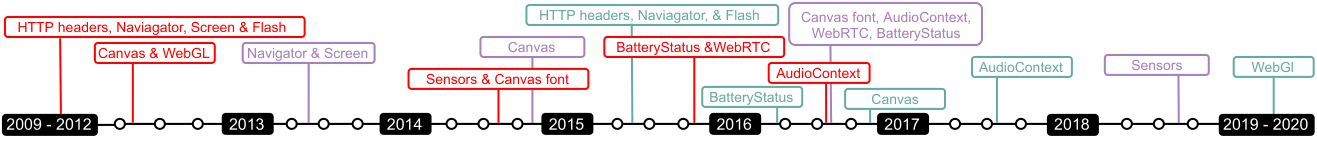}
  \caption{The timeline summarizes the chronological disclosure and adoption of fingerprinting APIs and countermeasures. Disclosures are represented with red, adoptions are represented with purple, and countermeasures are represented with green.}
\label{fig:apis-timeline}
\end{figure*}

\vspace{-.1in}
\subsection{Chronology of Browser Fingerprinting}
The fingerprinting surface has continued to expand with the introduction of new APIs and the disclosure of fingerprinting potential in existing APIs. 
Soon after an API is disclosed to have fingerprinting potential, they are adopted by trackers.
Countermeasures, also follow suit, and attempt to mitigate the fingerprintability of the API.
As shown in Figure \ref{fig:apis-timeline}, this pattern has been repeated over the years. 
We next provide a chronology of the disclosure, adoption, countermeasure of web APIs for fingerprinting.

\vspace{.05in} \noindent \textbf{Disclosure.}
Mayer \cite{mayer2009pamphleteer} first investigated browser fingerprinting in 2009 and showed that the fingerprints created through \texttt{navigator} and \texttt{screen} can uniquely identify 96.23\% of the browsers.
Soon after that in 2010, Eckersly \cite{eckersley2010unique} conducted a large scale user study to demonstrate that the information exposed through HTTP headers, e.g., \texttt{User-Agent} and APIs, e.g., \texttt{navigator}, and Flash, e.g. \texttt{fonts}, can be used to uniquely identify 94.2\% of the browsers.
In 2012, Mowery and Shacham \cite{mowery2012pixel} first introduced ``execution-based'' canvas and WebGL fingerprinting and showed that the certain images rendered through \texttt{canvas} and \texttt{WebGL} APIs on different devices produce different outputs due to the variance in hardware (e.g., graphics card) and software (e.g., browser version, configurations).
Since then researchers have demonstrated the fingerprinting potential of mobile sensors and canvas font in 2014 \cite{bojinov2014mobile, canvas_font_tor_bug}, \texttt{Battery Status} and \texttt{WebRTC} \cite{olejnik2015leaking,WebRTC_ip_leakage} in 2015, and \texttt{AudioContext} in 2016.

\vspace{.05in} \noindent \textbf{Adoption.}
Roughly after 2 years of disclosure, i.e., in 2013, browser fingerprinting, based on HTTP header information, JavaScript APIs, and Flash, was discovered on 40 of the top-10K websites \cite{nikiforakis2013cookieless}.
Within the next year, fingerprinting adoption exploded and canvas fingerprinting was discovered on 5,542 of top-100K websites, which is only 2 years after its initial disclosure \cite{Acar14WebNeverForgetsCCS}. 
The wide adoption of canvas fingerprinting was attributed to the release of fingerprintjs2 \cite{fingerprintjs2}, an open-source fingerprinting library. 
Later, in 2016, Englehardt et al., conducted a large scale study of top-1 million websites and further found the deployment of canvas font, \texttt{WebRTC}, \texttt{Audiocontext}, and \texttt{Battery Status} API fingerprinting on 14,371, 3,250, 715, 518, and 22 websites, respectively \cite{Englehardt16MillionSiteMeasurementCCS, Olejnik17BatteryStatusIWPE}, i.e., only after 1-2 years of their disclosure.
In 2018, Das et al. \cite{Das18MobileSensorsCCS} found the usage of sensors, such as motion and orientation, for browser fingerprinting on 3,695 of the top-100K websites, which is 4 years after their initial disclosure.

\vspace{.05in} \noindent \textbf{Countermeasures.}
Countermeasures against browser fingerprinting have a difficult time keeping up with the adoption of APIs for browser fingerprinting.
It nearly took 2 years, after the adoption of HTTP header (e.g., \texttt{User-Agent}) and APIs (e.g., \texttt{Navigator}) exploitation for fingerprinting to propose robust countermeasures against them \cite{Ferreira:2015, NikiforakisPriVaricator2015WWW}.
Similarly, the countermeasures against \texttt{Battery Status}, \texttt{canvas}, \texttt{AudioContext}, and \texttt{WebGL}, fingerprinting were first proposed in 2016 \cite{battery_status_removal}, 2016 \cite{baumann2016disguised}, 2017 \cite{laperdrix2017fprandom}, and 2019 \cite{Wu:2019}, respectively, which is nearly 1--7 years after their adoption. 
Some recent heuristics and machine learning approaches \cite{Englehardt16MillionSiteMeasurementCCS,iqbal21fpinspector, ddg_tracker_radar_fp,rizzo2021unveiling,Reitinger_2021_mlcb} have attempted to detect known fingerprinting techniques and block the scripts that implement them. 
Englehardt and Narayanan \cite{Englehardt16MillionSiteMeasurementCCS} proposed heuristics to detect fingerprinting scripts that implement \texttt{canvas}, \texttt{canvas Font}, and \texttt{webRTC} fingerprinting techniques.
They incidentally discovered the use of \texttt{AudioContext} fingerprinting in their manual analysis of the detected fingerprinting scripts. 
Iqbal et al. \cite{iqbal21fpinspector} proposed a supervised machine learning approach to detect fingerprinting scripts that implement various fingerprinting techniques, such as \texttt{canvas}, \texttt{canvas Font}, \texttt{webRTC}, \texttt{WebGL}, and \texttt{AudioContext}. 
They also incidentally discovered the potential use of peripheral probing (e.g., \texttt{getLayoutMap}) and \texttt{Permissions} API based fingerprinting in the post-hoc analysis of the detected fingerprinting scripts.
DuckDuckGo proposed to detect browser fingerprinting scripts based on the sum of ``API weights'' -- which are the ratio of API's appearance in ``suspicious scripts'' to ``non-suspicious scripts'' \cite{ddg_tracker_radar_fp}.
%
%
Based on the API weights, DuckDuckGo incidentally discovered the potential use of \texttt{deviceMemory} and \texttt{Presentation} APIs for browser fingerprinting \cite{ddg_fp_discoveries}. 

\vspace{-.1in}
\subsection{Takeaway}
\vspace{-.1in}
In conclusion, prior work is limited to reactive detection of scripts that implement known fingerprinting techniques.
Unsurprisingly, as discussed above, existing approaches have a difficult time keeping up because they are not designed to detect new fingerprinting techniques \cite{Englehardt16MillionSiteMeasurementCCS,iqbal21fpinspector,ddg_tracker_radar_fp}. 
Thus, as we discuss next, it is important to design approaches to detect new fingerprinting techniques in a timely fashion.

\section{\name}
\begin{figure*}[ht]
  \centering
  \includegraphics[width=\textwidth]{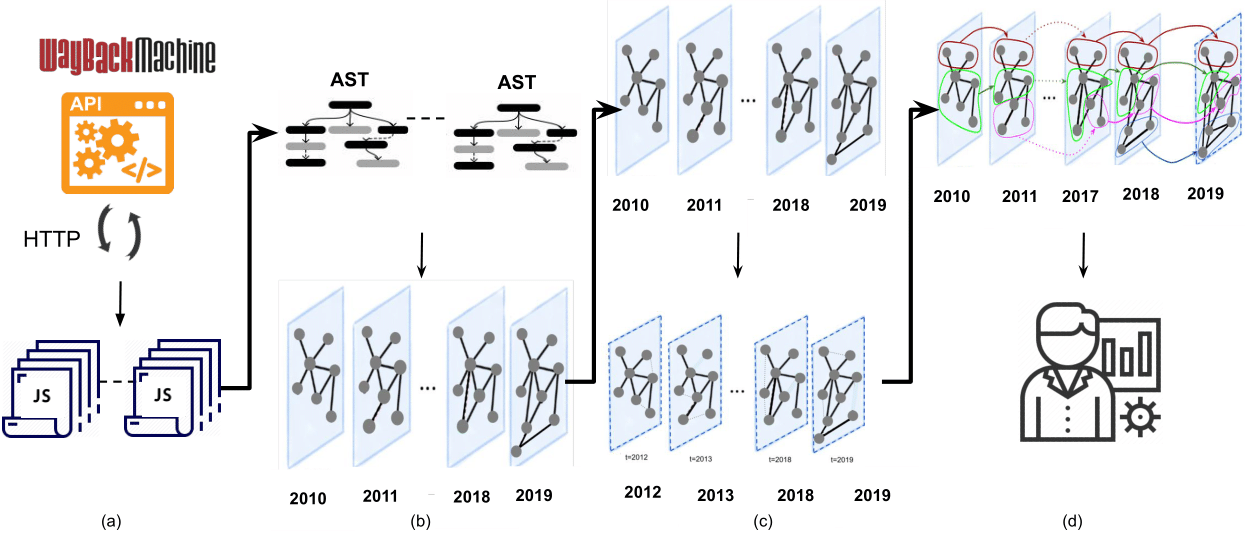}
  \caption{\name: (a) We use the Wayback Machine to crawl historical snapshots of scripts present on Alexa top-100K websites from 2010 to 2019, (b) We first create AST representation of crawled scripts to extract web API keywords, we then model the temporal co-occurrence of web APIs in scripts in a graph representation, (c) We leverages the temporal graph representation to predict future co-occurrence of web API keywords, (d) We first leverage the predicted co-occurrence to cluster web APIs based on their functionality, we then analyze the temporal clusters to detect abuse of specific API keywords for browser fingerprinting.}
\label{fig:fpgraph-overview}
\end{figure*}

We present the design and implementation of \name, a temporal graph based machine learning approach for early detection of web API abuse for browser fingerprinting.
As shown in Figure \ref{fig:fpgraph-overview}, \name can be divided into four components.
First, it models the temporal co-occurrence of web APIs in scripts using a graph representation. 
Second, it leverages the temporal graph representation to predict future co-occurrence of web APIs. 
Third, it leverages the predicted co-occurrence to cluster web APIs based on their functionality. 
Finally, the temporal clusters are analyzed to detect abuse of specific APIs (and their respective keywords) for browser fingerprinting.

\subsection{Modeling Temporal API Co-occurrence}
\WatchDog relies on the principle of \textit{guilt by association} to detect the abuse of web APIs for browser fingerprinting.
It means that if an API is being used alongside known fingerprinting APIs then we can presume that the API in question is also being abused for browser fingerprinting. 
We rely on the insight that trackers often use several fingerprinting techniques, and thus several fingerprinting APIs, together, to conduct browser fingerprinting \cite{Englehardt16MillionSiteMeasurementCCS,Das18MobileSensorsCCS,laperdrix2020browserfpsurvey,iqbal21fpinspector}.
\WatchDog operationalizes this insight in a longitudinal fashion to capture temporal trends in web API usage and early detection of web API abuse for browser fingerprinting.

\subsubsection{Longitudinal Data Crawling}
\label{section:longitudional-data-crawling}
To longitudinally analyze web APIs, \name needs to measure their usage on the web over time. 
We conduct a retrospective measurement study to analyze how web API usage has evolved on popular websites.
To gather historical snapshots of popular websites, we rely on the Internet Archive's Wayback Machine \cite{wayback}.
The Wayback Machine has periodically archived popular websites and their resources (e.g., scripts, images) since 1996 and has already archived more than 600 billion web pages thus far.
The Wayback Machine has been used in prior literature to conduct longitudinal measurements of online tracking \cite{Lerner16InternetJonesUSENIX,Iqbal17AntiABIMC}.

\vspace{.05in} \noindent \textbf{Crawling scripts using the Wayback Machine.}
\WatchDog relies on the Wayback Machine \cite{wayback} to crawl historical snapshots of a large set of scripts present on Alexa top-100K websites over the last decade (2010--2019).
Since crawling the Wayback Machine incurs significant additional overheads as compared to live web crawls, we limit our Wayback Machine crawls to scripts observed in our initial live crawl of Alexa top-10K websites and 10K websites randomly sampled from Alexa 10K-100K websites. 
To improve coverage of fingerprinting scripts, we further use the Wayback Machine to crawl historical snapshots of known fingerprinting scripts reported in recent prior work on Alexa top-100K websites \cite{iqbal21fpinspector}. 
It is noteworthy that \WatchDog is able to establish a comprehensive longitudinal view of web APIs usage because it conducts large-scale crawls of Alexa top-100K websites using Wayback machine instead of narrowly analyzing historical snapshots of a few number of JavaScript libraries for fingerprinting such as \texttt{fingerprintingjs2} \cite{fingerprintjs2}.

\vspace{.05in} \noindent \textbf{Completeness issues in the Wayback Machine.}
The Wayback Machine has completeness issues due to the inherent challenges of archiving the web \cite{HashimTakeMeBack07, KellyArchivability13, BrunelleNotAllMementos14, Lerner16InternetJonesUSENIX, Iqbal17AntiABIMC}.
First, the Wayback Machine used to not crawl websites based on their \texttt{robots.txt} policy.\footnote{Note that the Wayback Machine has resumed crawling websites since 2017 irrespective of their \texttt{robots.txt} policy \cite{internetarchive_policy}.}
Second, the Wayback Machine's crawls might miss dynamic resources. 
The Wayback Machine does not fully execute JavaScript during its archival process and thus misses some client-side dynamically generated URLs \cite{Lerner16InternetJonesUSENIX}.
Moreover, a resource might also not be captured by the Wayback Machine if the resource URL (file name or path) changes; thus the same resource is present with a different URL in the Wayback Machine's archival crawl as compared to our initial live crawl. 
Third, the Wayback Machine crawls less popular websites less frequently and thus might not crawl resources on low-ranked websites at least once every year.

\vspace{.05in} \noindent \textbf{Wayback Machine crawl statistics.}
Despite the aforementioned completeness issues in the Wayback Machine's crawls, we are able to longitudinally crawl yearly snapshots of almost 100K scripts from the Wayback Machine over the last decade (2010--2019).
Based on classification of \cite{fpinspector_CodeData}, this includes 1,658 fingerprinting and 92,193 non-fingerprinting scripts from our initial live crawl. 
Note that we use a two step process to crawl the Wayback Machine: we first fetch the URLs that point to the historical snapshots of scripts \cite{wayback_api} and then send requests for those URLs to gather their script content.
The first step returns URLs with the timestamp and the hash digest of the script content. 
The timestamps enables us to crawl scripts that are one year apart from each other and the hash digest helps us avoid crawling duplicate scripts in the second step.

We acknowledge that \WatchDog's longitudinal data collection misses a substantial number of scripts due to the completeness issues in the Wayback Machine. 
Specifically, with reference to our initial live crawl, we note that \name is unable to crawl snapshots of 43.09\% of the scripts from the Wayback Machine. 
While not ideal, we do not observe any bias in the missing scripts. 
Specifically, both fingerprinting and non-fingerprinting scripts are missed with roughly the same proportion, i.e., 43.60\% and 46.74\%, respectively. 
Moreover, despite the missing data, \WatchDog's longitudinal data collection is able to capture the overall trend of increasing adoption of browser fingerprinting over the years. 
Specifically, we observe fingerprinting scripts on 1.16\% and 3.70\% of the top-100K websites in 2016 and 2018, respectively.
This corroborates with the findings of prior studies of browser fingerprinting, which reported that 1.43\% of the top-million \cite{Englehardt16MillionSiteMeasurementCCS} and 3.69\% of the top-100K \cite{Das18MobileSensorsCCS} websites conduct browser fingerprinting in 2016 and 2018, respectively.
Thus, we conclude that \name's longitudinal data collection using the Wayback Machine is sufficient for us to retrospectively study the evolution of browser fingerprinting and draw meaningful conclusions.
We discuss alternates to the Wayback Machine and ideas to improve completeness of longitudinal crawls in Section \ref{sec: limitations}.

\subsubsection{Graph Representation}
\label{sec:guilt-association}
To model guilt by association, we represent web API co-occurrence in a graph.
Specifically, we model API keywords as nodes and include an edge between the nodes if the API keywords co-occur in the same script. 
We further weigh the edges based on the normalized co-occurrence frequency.

\vspace{.05in} \noindent \textbf{API keyword extraction.}
To extract API keywords from scripts, we model script text in abstract syntax trees (ASTs) that normalize scripts for developer coding styles.\footnote{We unpack \texttt{eval'ed} scripts with an instrumented browser \cite{Iqbal20AdGraphSP}. Unpacking allows us to treat scripts as code, which otherwise will be treated as a text string.}
ASTs also remove the non-essential script content (e.g., comments), generalize the APIs into generic primitives (e.g., \texttt{VariableDeclaration} and \texttt{ForStatement}), and capture the syntactical relationship between the APIs in form of a tree (e.g., an API call in a loop). 
Most importantly, ASTs provide a traverse-able tree representation of scripts, which allows us to extract the API keywords. 
We then traverse the ASTs from their roots to extract API keywords that match the standardized web APIs \cite{MDN_API}.

\begin{figure*}[ht]
  \centering
  \includegraphics[width=\linewidth]{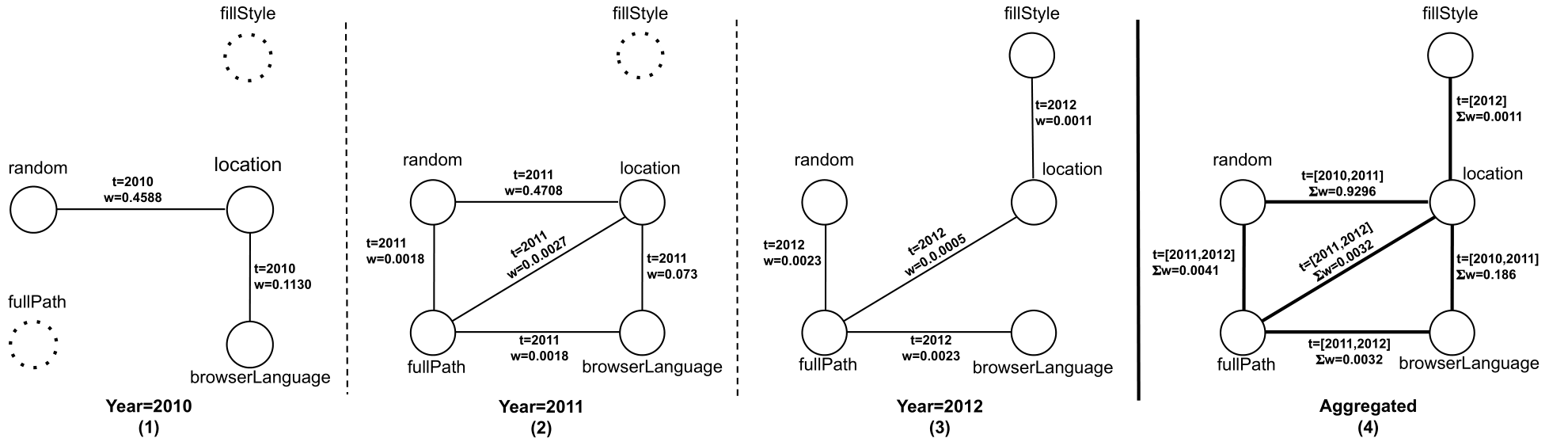}
  \caption{Creation of a sample temporal graph representation over years. (1), (2) and (3) represent the sub graphs for years 2010, 2011, and 2012, respectively. (4) represents the aggregated temporal graph.}
  \label{fig:temporal-graph-example}
\end{figure*}

\vspace{.05in} \noindent \textbf{Temporal graph representation.}
We capture the longitudinal co-occurrence of API keywords by annotating the edges with the timestamp, i.e., year, of API keyword co-occurrence. 
Furthermore, we capture the frequency of co-occurrence between the APIs over years by summing the edge weights. 
Figure \ref{fig:temporal-graph-example} demonstrates the creation of temporal graph. 
The figure shows a sample non-temporal graph representation for year 2010, 2011, and 2012 and their aggregated temporal graph representation. 
It can be seen in the aggregated graph representation that the edges are annotated with all the years in which the APIs co-occur and that the weight over the years is combined together.
For example, the weight of edge between \texttt{random} and \texttt{fullPath} in the aggregated graph (0.0041) is the sum of the weight of the edges between 2010 and 2012.

\subsection{Predicting API Co-occurrence}
To assist with \WatchDog's goal of early detection of web API abuse for browser fingerprinting, we attempt to predict API co-occurrence in the future. 
To this end, we leverage the longitudinal connectivity of APIs with each other to predict their future connectivity. 
We capture the longitudinal connectivity of APIs using \textit{hand-crafted} and \textit{graph-embedding} features. 
Our rationale for relying on these features is that the existing connectivity of APIs is indicative of their future connectivity.

\vspace{.05in} \noindent \textbf{Hand-crafted features.}
We first capture API co-occurrence patterns, targeting neighborhood connectivity, through hand-crafted features. 
These features model the connectivity between APIs, centrality of APIs, and the commonalities in API neighborhood. 
We also incorporate node weight and temporal information by giving more value to the recently formed edges. 
Specifically, the weight between two nodes is multiplied by a time factor, which decreases by one per year, for prior years. 
Incorporating weighted temporal information allows us to give more importance to the recent API co-occurrence patterns in the graph, which might be a better representative of the future connectivity between APIs.

We list hand-crafted features below:

\begin{enumerate}
    \item \textit{Common Neighbors:} The number of common neighbors between a node pair. The value is higher if the nodes have high number of common neighbors.
    
    \item \textit{Adamic-Adar Index:} The sum of the inverse logarithmic degree of the neighbors shared by a node pair. The nodes with fewer common neighbors have higher values. 
    
    \item \textit{Hub Promoted Index:} The number of common neighbors divided by the number of neighbors of the node with least degree in a node pair. The node pairs adjacent to hubs (high-degree nodes) have high values. 
    
    \item \textit{Hub Depressed Index:} The number of common neighbors divided by the number of neighbors of the node with highest degree in a node pair. The node pairs adjacent to hubs (high-degree nodes) have low values. 
    
    \item \textit{Jaccard Index:} The proportion of common neighbors by the total number of neighbors of a node pair. The value is higher if a node pair has more common neighbors in their neighborhood.
    
    \item \textit{Leicht-Holme-Newman Index:} The number of common neighbors divided by the product of the degree of the node pair. The value is higher if the nodes have low degree. 
    
    \item \textit{Resource Allocation Index:} The summation of the inverse of the degree of common neighbors between a node pair. The value is higher if the neighbors have low degree. 
    
    \item \textit{Salton Index (Cosine similarity):} It measures the cosine of the angle between the neighbors of a node pair. The more common the neighboring nodes, the higher the value. 
    
    \item \textit{Sorensen Similarity:} The proportion of the common neighbors by the sum of the degree of a node pair. The value is higher if the node pair has low degree. 
    
\end{enumerate}

\vspace{.05in}
Prior research \cite{butun2018extension} has shown that these features are highly predictive of the future connectivity in temporal graphs.
%
However, these features were only evaluated on temporal social network graphs and they may not be effective on temporal web API co-occurrence graphs.
To this end, we compute the information gain \cite{QuinlanInformationGain} (feature importance) of these features to evaluate their potential in predicting the future connectivity between web APIs in temporal API co-occurrence graphs.
Table \ref{tab:feature_importance} lists the information gain of hand-crafted features. 
It can be seen from the table that almost all feature provide an information gain of at least 5\% and the top three features provide the information gain of more than 12\%.
Overall, information gain indicates that the hand-crafted features are generic enough to be used for predicting future connectivity between web APIs in temporal API co-occurrence graphs. 

\begin{table}[h!]
\centering
\begin{tabular}{l|c}
\textbf{Features} & \textbf{Information gain (\%)} \\ 
\toprule
Leicht Holme Newman Index & 16.52 $\pm$ 1.63 \\
Temporal Edge Weight      & 13.42 $\pm$ 3.48 \\
Edge Weight               & 12.06 $\pm$ 3.16 \\
Salton                    & 8.61 $\pm$ 1.77 \\
Resource Allocation       & 8.5 $\pm$ 1.94 \\
Average Degree            & 7.8 $\pm$ 2.19 \\
Sorensen Similarity       & 6.43 $\pm$ 1.58 \\
Jaccard                   & 6 $\pm$ 1.6 \\
Common Neighbiors         & 5.74 $\pm$ 1.51 \\
Hub Depressed             & 5.5 $\pm$ 1.31 \\
Adamic-Adar               & 4.84 $\pm$ 1.19 \\
Hub Promoted              & 4.57 $\pm$ 2.07 \\
\end{tabular}
\caption{Hand-crafted features used by \name for graph prediction and information gain values (averaged over 10 years)}
\label{tab:feature_importance}
\end{table}

\textbf{Graph embedding-based features.}
\label{embeddings}
We capture more nuanced API co-occurrence patterns, potentially not modeled by our hand-crafted features, through graph embeddings. 
Graph embeddings encapsulate a node's neighborhood in a vector representation, such that the similar nodes in the graph have similar vector representation \cite{grover2016node2vec,perozzi2014deepwalk}. 
We determine a node's neighborhood through a series of biased random walks. 
Specifically, the random walks respect time order, i.e., edges are traversed in ascending order of time, and recently formed edges are selected with higher probability. 
Once a node's neighborhood is determined, it is mapped to an embedding space, such that the embeddings of two nodes that are similar to each other in the graph also have similar embeddings. 
After creating the node embeddings, we combine the emdeddings of a node pair using a weighted L2 regularization \cite{grover2016node2vec}.


\vspace{.05in} \noindent \textbf{Edge Prediction.}
We use a random forest \cite{Breiman01RandomForest} machine learning ensemble to predict the JavaScript APIs future co-occurrence.
Random forest combines the decisions from several decisions trees, each trained on a subset of features selected at random, and outputs the majority decision. 
We configure a random forest ensemble with 100 decision trees.

Each node in the decision tree is split using the best feature, based on information gain, among the subset of features. 
%
%
We note that our classes are imbalanced, i.e., API pairs are far less likely to not co-occur than they are to co-occur.
Thus, we bias our model by down sampling no-occurrence instances to the half of co-occurrence instances. 
Penalizing the model allows us to predict the APIs co-occurrence more favorably.

We predict the API co-occurrence over the year, i.e., from 2010--2020, by iteratively building the temporal graph. 
Specifically, as we move forward in time, our temporal graph contains APIs co-occurrences from all the snapshots thus far.
For example, for year 2010, our temporal graph only contains API co-occurrence that existed in year 2010, however, for year 2014, the temporal graph contains the API co-occurrence that existed between years 2010 and 2014.
For each year $Y$, we treat all possible API pairs, from the temporal graph of the last year $G_{Y-1}$, as probable candidates that may co-occur in the current year. 
The actual co-occurrence between the APIs in the current year $Y$, is considered as ground truth. 
We then use this information to train \name's random forest ensemble. 
Once we train the model, we use it to predict the future graph in the following year, i.e., year $Y+1$.
Specifically, we treat all possible API pairs, from the temporal graph of current year $G_{Y}$, as probable candidates that may co-occur next year. 
Since we are retrospectively predicting the APIs co-occurrence, we are in a unique position to also validate the predicted APIs co-occurrence that would happen in the future, i.e., by using the future API co-occurrence in year $Y+1$.
It is noteworthy that we combine both hand-crafted and graph embedding based features to train a combined random forest ensemble.

\begin{table*}[!htpb]
\centering
\begin{tabular}{c|l|l|ccc|ccc|ccc}
 \multirow{2}{*}{\textbf{Year}} & \textbf{\# of} & \textbf{\# of}  & \multicolumn{3}{c|}{\textbf{Hand-crafted}} & \multicolumn{3}{c|}{\textbf{Graph embeddings}} & \multicolumn{3}{c}{\textbf{Combined}} \\ \cline{4-12} 
 & \textbf{Nodes} & \textbf{Edges} & \textbf{Accuracy} & \textbf{Precision} & \textbf{Recall} & \textbf{Accuracy} & \textbf{Precision} & \textbf{Recall} & \textbf{Accuracy} & \textbf{Precision} & \textbf{Recall} \\ \hline

\textbf{2012} & 1,170 & 226,013 & 87.12 \% & 78.90\% & 71.10\%  & 79.40\% & 82.93\% & 86.99\% & 88.40\% & 89.50\% & 98.40\% \\ \hline
\textbf{2013} & 1,354 & 310,954 & 91.12\% & 84.25\% & 77.35\%  & 77.63\% & 79.16\% & 90.13\% & 86.50\% & 91.80\% & 93.10\% \\\hline
\textbf{2014} & 1.896 & 564,448 & 86.20\% & 89.40\% & 63.80\%  & 76.18\% & 71.93\% & 79.20\% & 90.10\% & 96.0\% & 93.40\% \\ \hline
\textbf{2015} & 2,096 & 746,379 & 86.50\% & 81.30\% & 72.20\%  & 71.93\% & 73.66\% & 90.76\% & 87.30\% & 89.70\% & 96.50\% \\ \hline
\textbf{2016} & 2,599 & 1,286,524 & 85.90\% & 82.10\% & 73.30\%  & 73.28\% & 75.16\% & 89.50\% & 87.70\% & 97.60\% & 89.20\% \\ \hline
\textbf{2017} & 2,978 & 1,669,328 & 87.30\% & 81.14\% & 83.30\%  & 74.75\% & 75.91\% & 91.01\% & 91.10\% & 94.80\% & 95.70\% \\ \hline
\textbf{2018} & 3,409 & 2,241,026 & 87.96\% & 80.90\% & 83.4\%  & 74.42\% & 75.87\% & 90.35\% & 90.30\% & 95.40\% & 94.20\% \\ \hline
\textbf{2019} & 3,603 & 2,684,157 & 88.41\% & 80.30\% & 81.20\%  & 78.14\% & 83.08\% & 84.41\% & 91.70\% & 92.70\% & 98.80\% \\ \midrule
\textbf{Mean} & 2,385 & 1,216,103 & 87.58\% & 82.29\% & 75.71\% & 75.71\% & 78.13\% & 90.08\% & 89.13\% & 93.44\% & 94.91\% \\
\end{tabular}
\caption{\name's accuracy in predicting APIs co-occurrence with hand-crafted and graph embedding-based features.}
\label{tab:results-table}
\end{table*}

\vspace{.05in} \noindent \textbf{Results.}
Table \ref{tab:results-table} presents \name's accuracy in predicting API co-occurrence over the years.
We provide separate as well as the combined accuracy of hand-crafted and graph embedding-based features.
Table \ref{tab:results-table} shows that \name's accuracy is significantly improved when hand-crafted and graph embedding-based features are combined together.
Specifically, the average accuracy, over the years, for hand-crafted features is {87.5\%} and graph-embedding based features is {76.24\%}.
When combined together, the mean accuracy increases to {88.13\%}.

\subsection{Clustering API Temporal Graphs}
\label{ZeroDay}
\name's temporal graphs allow us to longitudinally investigate the evolution of web API co-occurrence.
To this end, we partition temporal API graphs into clusters to systematically analyze APIs that are used for similar functionality together.
\name clusters the graphs based on the Louvain method \cite{blondel2008fast}, which partitions the graph such that the modularity is maximized between  clusters.
If a cluster contains more than one-third of the API keywords, \name partitions it again into sub-clusters.

\name clusters temporal API co-occurrence graphs and links clusters across consecutive years together to form temporal clusters. 
Specifically, \name links clusters together if their Jaccard similarity in more than 20\%.
If more than one cluster meets the similarity threshold in the prior year, they are merged together in the following year.
If a cluster from prior year matches more than one clusters in the following year, it is attached to all of the clusters in the following year.
If none of the clusters from prior years meet the similarity threshold, a new temporal cluster is created in the following year. 
Cluster from prior years that do not get attached to the clusters in the following year, may get attached to clusters in the coming years. 
Short-lived clusters, with a lifespan of at most 2 years, are filtered because they do not capture meaningful longitudinal trends.
Finally, \name extracts 14 temporal clusters.

\begin{figure}[ht]
  \centering
  \includegraphics[width=\linewidth]{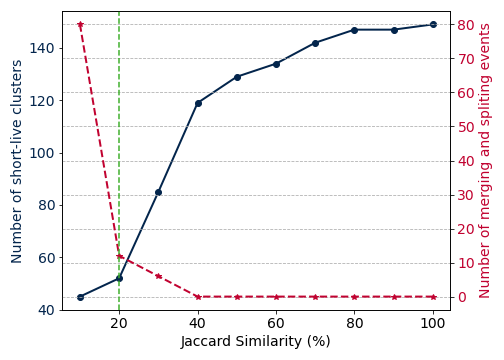}
  \caption{Number of short-lived temporal clusters along with number of merging and splitting events with different similarity thresholds.}
  \label{fig:similarity_justifivarion}
\end{figure}

\textbf{Jaccard similarity threshold.}
Figure \ref{fig:similarity_justifivarion} plots the trade-off between the number of short-lived clusters and the merging and splitting of clusters with varying Jaccard similarity threshold. 
It can be seen from the figure that as the similarity threshold increases, the number of short-lived clusters increases and the number of merging and splitting of clusters reduces. 
%
We pick 20\% as a threshold to link clusters, across consecutive years, because it provides the best trade-off. 
If we pick a higher similarity threshold, we risk losing a significant number of APIs, present in the short-lived clusters, and also risk merging clusters with varying functionality, together. 
%

%

\subsection{Labeling Temporal Clusters}
\label{sec:AnalysingTemporalClusters}
We next semi-automatically analyze the temporal clusters to label them. 
To this end, we expect functionally related APIs to appear together in a temporal cluster. 
To map keywords to their respective interfaces and APIs, we use MDN's \cite{MDN_API} hierarchical taxonomy of 88 APIs and 1024 interfaces.
We then identify the dominant APIs of each temporal cluster using this taxonomy. 
Specifically, we measure the dominance of an API in a cluster as the fraction of its keywords that exist in the cluster.

\vspace{.05in} \noindent \textbf{Labeling the fingerprinting cluster.}
Since different fingerprinting techniques are often used together \cite{Englehardt16MillionSiteMeasurementCCS,Das18MobileSensorsCCS,laperdrix2020browserfpsurvey,iqbal21fpinspector}, we expect that the web APIs abused for fingerprinting will be partitioned in a separate temporal cluster. 
To label the fingerprinting cluster, we analyze the following fingerprinting metrics for each of the 14 temporal clusters: 

\begin{enumerate}
    \item Percentage of API keywords that appear in fingerprinting scripts reported by \cite{iqbal21fpinspector}.
    \item Percentage of API keywords that are used in the open-source fingerprintjs2 fingerprinting library containing 152 API keywords \cite{fingerprintjs2}.
    \item Percentage of API keywords that \textit{only} appear in known fingerprinting scripts reported by \cite{iqbal21fpinspector} (i.e., not in any non-fingerprinting scripts).
    \item Ratio of the fraction of API keywords that appear in fingerprinting scripts to that in non-fingerprinting scripts as reported by \cite{iqbal21fpinspector}. 
    \end{enumerate}

\begin{table*}[!htpb]
\begin{tabular}{c|c|c|c|c|c|l}
\textbf{Cluster} & \textbf{Life-span} & \textbf{\% keywords} & \textbf{\% keywords } & \textbf{\% keywords in} & \textbf{FP/Non-} & \textbf{Dominant} \\
\textbf{size} & \textbf{in years} & \textbf{in FP scripts} & \textbf{in fpjs2 \cite{fingerprintjs2}} & \textbf{only FP scripts} & \textbf{FP ratio} & 
\textbf{APIs} \\ \hline
\cellcolor[HTML]{C62828}313 &  \cellcolor[HTML]{C62828}9 &  \cellcolor[HTML]{C62828}63\% & \cellcolor[HTML]{C62828}36\% & \cellcolor[HTML]{C62828}13\% & \cellcolor[HTML]{C62828}26.85 &  \cellcolor[HTML]{C62828}Battery, Navigator, Network Information \\ \hline
\cellcolor[HTML]{FFCDD2}{313} & \cellcolor[HTML]{FFCDD2}5  & \cellcolor[HTML]{FFCDD2}23\% & \cellcolor[HTML]{FFCDD2}2\%  & \cellcolor[HTML]{FFCDD2}0\% & \cellcolor[HTML]{FFCDD2}6.06 & \cellcolor[HTML]{FFCDD2}Long Tasks, Resource Timing, Background Tasks  \\ \hline
{256} & 4  & 15\% & 1\%  & 0 & 1.43 & XMLSerializer, Mouse, ShadowRoot \\ \hline
\cellcolor[HTML]{FFCDD2}{222} & \cellcolor[HTML]{FFCDD2}6  & \cellcolor[HTML]{FFCDD2}17\% & \cellcolor[HTML]{FFCDD2}6\% & \cellcolor[HTML]{FFCDD2}2\% & \cellcolor[HTML]{FFCDD2}3.13 & \cellcolor[HTML]{FFCDD2}Mouse, TouchEvents, Canvas \\ \hline
{161} & 4  & 11\% & 3\% & 2\% & 0.89 & CSS Painting, XMLHTTPRequest \\ \hline
{143} & 4  & 11\% & 3\% & 0 & 0.79 & VideoTrack, Geolocation, Long Tasks\\ \hline
\cellcolor[HTML]{FFCDD2}{142} & \cellcolor[HTML]{FFCDD2}10 & \cellcolor[HTML]{FFCDD2}20\% & \cellcolor[HTML]{FFCDD2}5\% & \cellcolor[HTML]{FFCDD2}0 & \cellcolor[HTML]{FFCDD2}0.61 & \cellcolor[HTML]{FFCDD2}HTMLIFrameElement, Navigator, URL \\ \hline
{141} & 5  & 15\% & 4\% & 0 & 1.19 & Visual Viewport, Crypto, Channel Messaging \\ \hline
{125} & 4  & 10\% & 2\% & 0 & 0.25   & Fetch API, Notification, NodeFilter \\ \hline
\cellcolor[HTML]{FFCDD2}{121} & \cellcolor[HTML]{FFCDD2}9  & \cellcolor[HTML]{FFCDD2}18\% & \cellcolor[HTML]{FFCDD2}4\% & \cellcolor[HTML]{FFCDD2}0\% & \cellcolor[HTML]{FFCDD2}1.24 & \cellcolor[HTML]{FFCDD2}Resource Timing, Page Visibility, History \\ \hline
\cellcolor[HTML]{FFCDD2}{92}  & \cellcolor[HTML]{FFCDD2}5  & \cellcolor[HTML]{FFCDD2}20\% & \cellcolor[HTML]{FFCDD2}2\% & \cellcolor[HTML]{FFCDD2}0\% & \cellcolor[HTML]{FFCDD2}0.98 & \cellcolor[HTML]{FFCDD2}FullScreen, VideoTrack, HTMLMediaElement \\ \hline
{78}  & 4  & 3\%  & \textless{}1\% & 3\% & 1.94 & FileReader, Web Animations, XMLHttpRequest \\ \hline
{67}  & 4  & 3\%  & 2\% & 0 & 0.27   & Sensors, Gamepad, Fullscreen, Web Bluetooth \\ \hline
{28}  & 4  & 2\%  & 1\% & 0 & 0.00   & History, HTMLElement, HTMLTableElement \\     
\end{tabular}

\caption{Temporal clusters detected by \name and their key characteristics. Based on their fingerprinting potential, clusters are marked with different gradients of red. The fingerprinting cluster (represented by \colorbox[HTML]{C62828}{dark red}) clearly stands out as compared to the remaining clusters in terms of its similarity with known fingerprinting scripts.}
\label{tab:temporal-clusters}
\end{table*}

Note that \name partially relies on FP-Inspector \cite{iqbal21fpinspector} to label the fingerprinting cluster. 
However, we argue that it is the best available ground truth for browser fingerprinting, as compared to using other alternatives such as filter lists. 
Disconnect \cite{disconnect_me} only provides the domain names of fingerprinting vendors, rather than the full URLs of fingerprinting scripts, and thus cannot distinguish between fingerprinting and non-fingerprinting resources served from the same domain. 


\vspace{.05in} \noindent \textbf{Results.}
Table \ref{tab:temporal-clusters} shows the temporal clusters and their key characteristics. 
Each row represents a cluster and the rows are sorted based on the cluster size. 
We note that the top-ranked cluster clearly has significantly more pronounced fingerprinting metrics than other clusters, we label it as fingerprinting and the remaining as other.
First, 63\% of the keywords in the fingerprinting cluster are used in fingerprinting scripts, which is at least $\approx$3X more than any other temporal cluster.
Second, 36\% of the keywords in the fingerprinting cluster are used in fingerprintjs2, which is at least $\approx$6X more than any other temporal cluster. 
Third, 13\% of the keywords exclusively appear in fingerprinting scripts, which is at least $\approx$4X more than any other temporal cluster. 
Finally, the fraction of the keywords appearance in fingerprinting to non-fingerprinting scripts is 26.85, which is $\approx$4X more than any other temporal cluster.


\section{Analysis of APIs in the Fingerprinting Cluster}
\label{sec: ResultsDiscussion}
In this section, we conduct an in-depth analysis of the fingerprinting cluster detected by \name. 
Table \ref{tab: APIs} lists a subset of the keywords of top dominant APIs in the fingerprinting cluster.\footnote{We select a representative subset out of 313 total keywords to capture diverse use cases and cover almost all of the time-to-detection categories for each API. We will include the complete table along with the code/data as part of the artifact release. } 
We investigate how the functionality of dominant APIs is being abused for fingerprinting. 
We also assess the time-to-detection of \name as compared to their browser release and disclosure dates.
For each API keyword, we define release, disclosure, and detection dates as follows:

\begin{enumerate}
    \item \textit{Release} refers to the earliest date of support by one of the major browsers (i.e., Chrome, Firefox, Safari).
    \item \textit{Appearance} refers to the earliest date when the API keyword appeared in our dataset.
    \item  \textit{Disclosure} refers to the earliest date that a proof-of-concept fingerprinting design or implementation involving the API keyword was presented in a research publication, W3C documentation, or public forums.
    \item  \textit{Detection} refers to the earliest date when the API keyword was detected as a member of the fingerprinting cluster by \name.
\end{enumerate}

Based on this information, we classify each API keyword in the fingerprinting cluster into the following 4 categories:

\begin{enumerate}

\item {\name detects abuse of API-keywords that are yet \textbf{undisclosed} to the best of our knowledge.}
Denoted with {\colorbox[HTML]{82C785}{green}} color in Table \ref{tab: APIs}, \name detects a number of yet-undislosed API keywords such as \texttt{deviceMemeory} (Navigator), \texttt{WebGL2RenderingContext} (WebGL), \texttt{illuminance} (Sensor), and \texttt{paint} (Performance).

\item {\name detects abuse of APIs \textbf{before disclosure}.}
Denoted with {\colorbox[HTML]{FFEB3B}{yellow}} color in Table \ref{tab: APIs}, \name detects a number of  API keywords before their disclosure such as \texttt{getGamepads} (GamePad), \texttt{visibilityState} (Page Visibility), and \texttt{clipboardData} (Clipboard).

\item {\name detects abuse of APIs \textbf{after disclosure}.}
Denoted with {\colorbox[HTML]{F77B72}{red}} color in Table \ref{tab: APIs}, \name detects some API keywords after their disclosure such as \texttt{longitude} (Geolocation), \texttt{DeviceMotionEvent} (Sensor), and \texttt{plugins} (Navigator).

\item {\name detects abuse of APIs \textbf{at disclosure}.}
Denoted with {\colorbox[HTML]{FFC166}{orange}} color in Table \ref{tab: APIs}, \name detects a number of API keywords at their disclosure such as \texttt{chargingTime} (Battery Status), \texttt{now} (Performance), and \texttt{force} (Touch). 
Note that most of the late detections are in fact detected as early as possible by \name because the API keywords did not appear in our data before the detection date. 
In other words, \name detects these API keywords at the first possible opportunity. %
We also denote these with orange color in Table \ref{tab: APIs} and include API keywords such as \texttt{altitudeAccuracy} (Geolocation), \texttt{bufferData} (WebGL), and \texttt{chargingchange} (Battery Status).

\end{enumerate}



\begin{table*}
  \begin{tabular}[c]{c|l|c|c|c|c}
     \textbf{API Name} & \textbf{Keywords} & \textbf{Release Date} & \textbf{Appearance Date} & \textbf{Disclosure Date} & \textbf{Detection Date} \\
     
    \midrule
    \multirow{4}{*}{\textbf{Battery Status}}&
    \cellcolor[HTML]{FFC166} chargingTime & 
    \cellcolor[HTML]{FFC166} 2014 & 
    \cellcolor[HTML]{FFC166} 2015 & 
    \cellcolor[HTML]{FFC166} 2015 \cite{olejnik2015leaking} & 
    \cellcolor[HTML]{FFC166} 2015 \\ 
  & 
  \cellcolor[HTML]{FFC166} chargingchange &
  \cellcolor[HTML]{FFC166} 2014 &
  \cellcolor[HTML]{FFC166} 2017 & 
  \cellcolor[HTML]{FFC166} 2015 \cite{olejnik2015leaking} &
  \cellcolor[HTML]{FFC166} 2017 \\
  & 
  \cellcolor[HTML]{FFC166} dischargingTime &
  \cellcolor[HTML]{FFC166} 2014 & 
  \cellcolor[HTML]{FFC166} 2016 & 
  \cellcolor[HTML]{FFC166} 2015 \cite{olejnik2015leaking} & 
  \cellcolor[HTML]{FFC166} 2016 \\
  \midrule
  
  \multirow{7}{*}{\textbf{Navigator}} &
  \cellcolor[HTML]{82C785}deviceMemory &
  \cellcolor[HTML]{82C785}2017 &
  \cellcolor[HTML]{82C785}2017&
  \cellcolor[HTML]{82C785}- &
  \cellcolor[HTML]{82C785}2019 \\
 &
  \cellcolor[HTML]{FFEB3B}hardwareConcurrency &
  \cellcolor[HTML]{FFEB3B}2014 &
  \cellcolor[HTML]{FFEB3B}2014&
  \cellcolor[HTML]{FFEB3B}2017 \cite{saito2017web} &
  \cellcolor[HTML]{FFEB3B}2014 \\
 &
  \cellcolor[HTML]{FFC166}oscpu &
  \cellcolor[HTML]{FFC166}2004 &
  \cellcolor[HTML]{FFC166}2015&
  \cellcolor[HTML]{FFC166}2009 \cite{mayer2009pamphleteer} &
  \cellcolor[HTML]{FFC166}2015 \\
 &
  \cellcolor[HTML]{F77B72}plugins &
  \cellcolor[HTML]{F77B72}2003 &
  \cellcolor[HTML]{F77B72}2011&
  \cellcolor[HTML]{F77B72}2009 \cite{mayer2009pamphleteer} &
  \cellcolor[HTML]{F77B72}2016 \\
&
  \cellcolor[HTML]{82C785}vendorSub &
  \cellcolor[HTML]{82C785}2004 &
  \cellcolor[HTML]{82C785} 2013&
  \cellcolor[HTML]{82C785}- &
  \cellcolor[HTML]{82C785}2013 \\
  &
  \cellcolor[HTML]{82C785}webdriver &
  \cellcolor[HTML]{82C785}2015 &
  \cellcolor[HTML]{82C785}2014 &
  \cellcolor[HTML]{82C785}- &
  \cellcolor[HTML]{82C785}2017 \\

  \midrule
  \multirow{4}{*}{\textbf{Network Information}} &
  \cellcolor[HTML]{FFEB3B}downlink &
  \cellcolor[HTML]{FFEB3B}2017 &
  \cellcolor[HTML]{FFEB3B}2018&
  \cellcolor[HTML]{FFEB3B}2020 \cite{the_network_information_api_2020} &
  \cellcolor[HTML]{FFEB3B}2018 \\
 &
  \cellcolor[HTML]{FFEB3B}downlinkMax &
  \cellcolor[HTML]{FFEB3B}2017 &
  \cellcolor[HTML]{FFEB3B} 2018&
  \cellcolor[HTML]{FFEB3B}2020 \cite{the_network_information_api_2020} &
  \cellcolor[HTML]{FFEB3B}2019 \\
 &
  \cellcolor[HTML]{FFEB3B}rtt&
  \cellcolor[HTML]{FFEB3B}2017 &
  \cellcolor[HTML]{FFEB3B} 2017&
  \cellcolor[HTML]{FFEB3B}2020 \cite{the_network_information_api_2020} &
  \cellcolor[HTML]{FFEB3B}2017 \\

  \midrule
  \multirow{4}{*}{\textbf{Geolocation}} &
  \cellcolor[HTML]{FFC166}altitudeAccuracy &
  \cellcolor[HTML]{FFC166}2009 &
  \cellcolor[HTML]{FFC166} 2016  &
  \cellcolor[HTML]{FFC166}2008 \cite{geolocation_standardization_2008}&
  \cellcolor[HTML]{FFC166}2016 \\
 &
 \cellcolor[HTML]{FFC166}geolocation &
  \cellcolor[HTML]{FFC166}2009 &
  \cellcolor[HTML]{FFC166}2012  &
  \cellcolor[HTML]{FFC166}2008 \cite{geolocation_standardization_2008} &
  \cellcolor[HTML]{FFC166}2011 \\
 &
  \cellcolor[HTML]{F77B72}longitude &
  \cellcolor[HTML]{F77B72}2009 &
  \cellcolor[HTML]{F77B72}2012 &
  \cellcolor[HTML]{F77B72}2008 \cite{geolocation_standardization_2008} &
  \cellcolor[HTML]{F77B72}2016 \\
 &
  \cellcolor[HTML]{F77B72}watchPosition &
  \cellcolor[HTML]{F77B72}2009 &
  \cellcolor[HTML]{F77B72}2014 &
  \cellcolor[HTML]{F77B72}2008 \cite{geolocation_standardization_2008}&
  \cellcolor[HTML]{F77B72}2019 \\
  
  \midrule
  \multirow{7}{*}{\textbf{WebGL}} &
  \cellcolor[HTML]{FFC166}bufferData &
  \cellcolor[HTML]{FFC166}2011 &
  \cellcolor[HTML]{FFC166}2014&
  \cellcolor[HTML]{FFC166}2012 \cite{mowery2012pixel}&
  \cellcolor[HTML]{FFC166}2014 \\
  
 &
  \cellcolor[HTML]{F77B72}viewport &
  \cellcolor[HTML]{F77B72}2011 &
  \cellcolor[HTML]{F77B72}2012 &
  \cellcolor[HTML]{F77B72}2012 \cite{mowery2012pixel} &
  \cellcolor[HTML]{F77B72}2018 \\
 &
  \cellcolor[HTML]{FFEB3B}webgl &
  \cellcolor[HTML]{FFEB3B}2011 &
  \cellcolor[HTML]{FFEB3B}2011 &
  \cellcolor[HTML]{FFEB3B}2012 \cite{mowery2012pixel}&
  \cellcolor[HTML]{FFEB3B}2011 \\
 &
  \cellcolor[HTML]{FFEB3B}WEBGL\_debug\_renderer\_info &
  \cellcolor[HTML]{FFEB3B}2014 &
  \cellcolor[HTML]{FFEB3B}2014 &
  \cellcolor[HTML]{FFEB3B}2016 \cite{Englehardt16MillionSiteMeasurementCCS} &
  \cellcolor[HTML]{FFEB3B}2014 \\
  &
  \cellcolor[HTML]{82C785}WEBGL\_depth\_texture &
  \cellcolor[HTML]{82C785}2013 &
  \cellcolor[HTML]{82C785}2017 &
  \cellcolor[HTML]{82C785}- &
  \cellcolor[HTML]{82C785}2017 \\
  &
  \cellcolor[HTML]{82C785}WebGL2RenderingContext &
  \cellcolor[HTML]{82C785}2017 &
  \cellcolor[HTML]{82C785}2017 &
  \cellcolor[HTML]{82C785}- &
  \cellcolor[HTML]{82C785}2017 \\
  
  \midrule 
  \multirow{4}{*}{\textbf{Performance}}&
  \cellcolor[HTML]{82C785}domainLookupEnd &
  \cellcolor[HTML]{82C785}2015 &
  \cellcolor[HTML]{82C785}2015 &
  \cellcolor[HTML]{82C785}- &
  \cellcolor[HTML]{82C785}2016 \\
 &
  \cellcolor[HTML]{82C785}domainLookupStart &
  \cellcolor[HTML]{82C785}2015 &
  \cellcolor[HTML]{82C785}2015 &
  \cellcolor[HTML]{82C785}- &
  \cellcolor[HTML]{82C785}2016 \\
 &
  \cellcolor[HTML]{FFC166}now &
  \cellcolor[HTML]{FFC166}2012 &
  \cellcolor[HTML]{FFC166}2012 &
  \cellcolor[HTML]{FFC166}2016 \cite{performance2016} &
  \cellcolor[HTML]{FFC166}2016 \\
 &
  \cellcolor[HTML]{82C785}paint &
  \cellcolor[HTML]{82C785}2017 &
  \cellcolor[HTML]{82C785}2018 &
  \cellcolor[HTML]{82C785}- &
  \cellcolor[HTML]{82C785}2019 \\

  \midrule
  \multirow{4}{*}\textbf{\textbf{Page Visibility}} &
  \cellcolor[HTML]{FFEB3B}visibilityState &
  \cellcolor[HTML]{FFEB3B}2013 &
  \cellcolor[HTML]{FFEB3B}2013 &
  \cellcolor[HTML]{FFEB3B}2020 \cite{Ephemeralpagevisibility2020}&
  \cellcolor[HTML]{FFEB3B}2017 \\
  &
  \cellcolor[HTML]{FFEB3B}focused &
  \cellcolor[HTML]{FFEB3B}2013 &
  \cellcolor[HTML]{FFEB3B}2013 &
  \cellcolor[HTML]{FFEB3B}2020 \cite{Ephemeralpagevisibility2020} &
  \cellcolor[HTML]{FFEB3B}2013 \\
 &
  \cellcolor[HTML]{82C785}prerender &
  \cellcolor[HTML]{82C785}2013 &
  \cellcolor[HTML]{82C785}2013 &
  \cellcolor[HTML]{82C785}- &
  \cellcolor[HTML]{82C785}2016 \\

  \midrule
 \textbf{Web Worker} &
  \cellcolor[HTML]{FFEB3B}applicationCache &
  \cellcolor[HTML]{FFEB3B}2010 &
  \cellcolor[HTML]{FFEB3B}2011 &
  \cellcolor[HTML]{FFEB3B}2017 \cite{full-third-party-cookie-blocking} &
  \cellcolor[HTML]{FFEB3B}2011 \\
  &
  \cellcolor[HTML]{82C785}Worklet &
  \cellcolor[HTML]{82C785}2018 &
  \cellcolor[HTML]{82C785}2018 &
  \cellcolor[HTML]{82C785} - &
  \cellcolor[HTML]{82C785}2018 \\

  \midrule 
  \multirow{3}{*}{\textbf{GamePad}} &
  \cellcolor[HTML]{FFEB3B}Gamepad &
  \cellcolor[HTML]{FFEB3B}2014 &
  \cellcolor[HTML]{FFEB3B}2014 &
  \cellcolor[HTML]{FFEB3B}2020 \cite{securinggamepad2020} &
  \cellcolor[HTML]{FFEB3B}2019 \\
 &
  \cellcolor[HTML]{FFEB3B}getGamepads &
  \cellcolor[HTML]{FFEB3B}2014 &
  \cellcolor[HTML]{FFEB3B}2014 &
  \cellcolor[HTML]{FFEB3B}2020 \cite{securinggamepad2020} &
  \cellcolor[HTML]{FFEB3B}2014 \\
 &
  \cellcolor[HTML]{82C785}mapping &
  \cellcolor[HTML]{82C785}2014 &
  \cellcolor[HTML]{82C785}2015 &
  \cellcolor[HTML]{82C785}- &
  \cellcolor[HTML]{82C785}2017 \\

  \midrule
  {\textbf{Mouse}} & 
   \cellcolor[HTML]{FFC166}movementX &
   \cellcolor[HTML]{FFC166}2014 &
   \cellcolor[HTML]{FFC166}2016 &
   \cellcolor[HTML]{FFC166}2013 \cite{shahzad2013secure} &
   \cellcolor[HTML]{FFC166}2016 \\
  &
  \cellcolor[HTML]{F77B72}onmousemove &
  \cellcolor[HTML]{F77B72}2003 &
  \cellcolor[HTML]{F77B72}2012 &
  \cellcolor[HTML]{F77B72}2004 \cite{pusara2004user} &
  \cellcolor[HTML]{F77B72}2018 \\
  
  \midrule
  \multirow{4}{*}{\textbf{Touch}} &
   \cellcolor[HTML]{FFC166}force &
  \cellcolor[HTML]{FFC166}2012 &
  \cellcolor[HTML]{FFC166}2013 &
  \cellcolor[HTML]{FFC166}2013 \cite{shahzad2013secure} &
  \cellcolor[HTML]{FFC166}2013 \\
 &
  \cellcolor[HTML]{FFEB3B}ontouchstart &
  \cellcolor[HTML]{FFEB3B}2011 &
  \cellcolor[HTML]{FFEB3B}2011 &
  \cellcolor[HTML]{FFEB3B}2013 \cite{shahzad2013secure} &
  \cellcolor[HTML]{FFEB3B}2011 \\
&
  \cellcolor[HTML]{82C785}rotationAngle &
  \cellcolor[HTML]{82C785}2015 &
  \cellcolor[HTML]{82C785}2017 &
  \cellcolor[HTML]{82C785}- &
  \cellcolor[HTML]{82C785}2017 \\
 &
  \cellcolor[HTML]{F77B72}touchenter &
  \cellcolor[HTML]{F77B72}2012 &
  \cellcolor[HTML]{F77B72}2015 &
  \cellcolor[HTML]{F77B72}2013 \cite{shahzad2013secure} &
  \cellcolor[HTML]{F77B72}2016 \\
  
  \midrule
 \multirow{4}{*}{\textbf{Sensor}} & 
  \cellcolor[HTML]{82C785}AbsoluteOrientationSensor &
  \cellcolor[HTML]{82C785}2018 &
  \cellcolor[HTML]{82C785}2018 &
  \cellcolor[HTML]{82C785}- &
  \cellcolor[HTML]{82C785}2018 \\
 &
  \cellcolor[HTML]{82C785}AmbientLightSensor &
  \cellcolor[HTML]{82C785}2017 &
  \cellcolor[HTML]{82C785}2017 &
  \cellcolor[HTML]{82C785}- &
  \cellcolor[HTML]{82C785}2017 \\
 &
  \cellcolor[HTML]{F77B72}acceleration &
  \cellcolor[HTML]{F77B72}2011 &
  \cellcolor[HTML]{F77B72}2013 &
  \cellcolor[HTML]{F77B72}2014 \cite{bojinov2014mobile} &
  \cellcolor[HTML]{F77B72}2017 \\
 &
  \cellcolor[HTML]{F77B72}DeviceMotionEvent &
  \cellcolor[HTML]{F77B72}2014 &
  \cellcolor[HTML]{F77B72}2013 &
  \cellcolor[HTML]{F77B72}2014 \cite{bojinov2014mobile} &
  \cellcolor[HTML]{F77B72}2018 \\
 &
 \cellcolor[HTML]{82C785}illuminance &
  \cellcolor[HTML]{82C785}2017 &
  \cellcolor[HTML]{82C785}2017 &
  \cellcolor[HTML]{82C785}- &
  \cellcolor[HTML]{82C785}2018 \\
 &
  \cellcolor[HTML]{82C785}Magnetometer &
  \cellcolor[HTML]{82C785}2017 &
  \cellcolor[HTML]{82C785}2017 &
  \cellcolor[HTML]{82C785}- &
  \cellcolor[HTML]{82C785}2018 \\
 &
  \cellcolor[HTML]{82C785}rotationRate &
  \cellcolor[HTML]{82C785}2011 &
  \cellcolor[HTML]{82C785}2017 &
  \cellcolor[HTML]{82C785}- &
  \cellcolor[HTML]{82C785}2018 \\
  
  \midrule
  
\multirow{3}{*}{\textbf{Clipboard}}&
\cellcolor[HTML]{FFEB3B}copy &
  \cellcolor[HTML]{FFEB3B}2007 &
  \cellcolor[HTML]{FFEB3B}2018 &
  \cellcolor[HTML]{FFEB3B}2020 \cite{AsyncClipboard}&
  \cellcolor[HTML]{FFEB3B}2019 \\
 &
  \cellcolor[HTML]{FFEB3B}clipboardData &
  \cellcolor[HTML]{FFEB3B}2013 &
  \cellcolor[HTML]{FFEB3B}2018 &
  \cellcolor[HTML]{FFEB3B}2020 \cite{AsyncClipboard}&
  \cellcolor[HTML]{FFEB3B}2018 \\
 &
  \cellcolor[HTML]{FFEB3B}paste &
  \cellcolor[HTML]{FFEB3B}2007 &
  \cellcolor[HTML]{FFEB3B}2018 &
  \cellcolor[HTML]{FFEB3B}2020 \cite{AsyncClipboard}&
  \cellcolor[HTML]{FFEB3B}2019 \\
     \end{tabular} 
     \caption{List of dominant API detected by \name and their time-to-detection:  {\colorbox[HTML]{82C785}{not-yet-disclosed}} {\colorbox[HTML]{FFEB3B}{early detection}} {\colorbox[HTML]{FFC166}{on-time detection}} {\colorbox[HTML]{F77B72}{late detection}}.}
  \label{tab: APIs}
\end{table*}


Next, we do a manual deep dive into each of the APIs listed in Table \ref{tab: APIs} in the descending order of their dominance.
%
%
Note that we do not discuss some of the known web APIs, such as \texttt{canvas} and \texttt{canvas font}, \texttt{webRTC}, \texttt{AudioContext}, that are already shown to be widely abused for browser fingerprinting \cite{Englehardt16MillionSiteMeasurementCCS,Acar14WebNeverForgetsCCS}. 

\textbf{\texttt{Battery Status}}, standardized in 2011 \cite{battery_status_standardization_2011} and supported in major browsers as early as 2014, is a non-permissioned API that provides information about a device's battery status to help web applications adjust resource usage when battery power is low.
In 2015, Olejnik et al. disclosed that battery capacity and charging level can be abused for fingerprinting \cite{olejnik2015leaking}.\footnote{Due to these fingerprinting concerns \cite{RemovingBattery2016}, Firefox stopped supporting the API in 2017 \cite{removebatterybugzilla2016}.}
More specifically, the information about current battery level (\texttt{level}) and predicted time to charge (\texttt{chargingTime}) or discharge (\texttt{dischargingTime}) can be used to estimate a device's battery capacity, which is lower than its design capacity and often distinctive. 
\WatchDog detects these keywords in 2015, right at the time of disclosure. %
Furthermore, \WatchDog detects a change in the abuse of \texttt{Battery Status} API staring 2017.
More specifically, fingerprinters started gathering the change frequency of the battery status using keywords such as \texttt{chargingchange}, \texttt{chargingtimechange}, \texttt{dischargingtimechange}, and \texttt{levelchange} that reflect different workloads to create short-lived fingerprint.
Script \ref{script:battery_status} shows a fingerprinting snippet that uses the aforementioned keywords. 

\textbf{\texttt{Navigator}}, standardized in 1997 and supported by all major browsers since then \cite{mdn-navigator}, is a non-permissioned interface that provides information about the browser.
In 2009, Mayer \cite{mayer2009pamphleteer} disclosed that the navigator object provides information about a browser's settings that can be abused for fingerprinting. 
More specifically, the user agent string (\texttt{userAgent}), the languages supported by the browser (\texttt{languages}), the list of installed plugins (\texttt{plugins}), and supported file formats (\texttt{mimeType}) can reveal distinctive information about a browser.
Since these features individually might not be sufficient to uniquely identify a browser, fingerprinters tend to gather a number of device-specific information exposed by navigator to increase the entropy of the fingerprint \cite{laperdrix2020browserfpsurvey}.
Script \ref{script:navigator} shows a fingerprinting snippet that gathers 18 different navigator properties including the aforementioned keywords. 
\WatchDog detects navigator-related keywords as early as 2013, which is roughly around the time when researchers first documented fingerprinting on the web through large-scale measurements \cite{nikiforakis2013cookieless}.
%
Note that the \texttt{Navigator} interface has been updated several times over the years to support new features. 
\WatchDog is able to detect the abuse of most of the newly introduced navigator properties in a timely fashion. 
For example, \WatchDog detects \texttt{hardwareConcurrency}, which returns the available number of logical processor cores, in 2014 right after its standardization even though its abuse was disclosed later in 2017 \cite{saito2017web}.

\textbf{\texttt{Network Information}} API, standardized in 2014 \cite{network_information_standardization_2014} and supported by major mobile browsers (except Safari) since 2017 \cite{Network_Information_API}, is a non-permissioned API that provides network connection information of the browser.
More specifically, connection type (\texttt{type}, such as WiFi, WiMAX, Ethernet), delay (\texttt{rtt}), bandwidth (\texttt{downlink} and \texttt{downlinkMax}), and change in connection type (\texttt{onchange}) information are accessible via this API.
It is noteworthy that potential privacy issues of the Network Status API were originally dismissed by W3C (``minimal impact on privacy or fingerprinting'')  \cite{the_network_information_api_2012} and none of the prior fingerprinting measurement studies report its abuse \cite{mayer2009pamphleteer, eckersley2010unique, mowery2012pixel, canvas_font_tor_bug, olejnik2015leaking, WebRTC_ip_leakage}.
However, as later acknowledged by W3C in 2020 \cite{the_network_information_api_2020}, 
this information could be abused to fingerprint a user based on the time and order of transitions between networks as well as user location.
Note that Firefox and Safari explicitly declined to support this API due to fingerprinting concerns \cite{bugzilla_networkinformation, apple_declined_16_apis}.
%
%
\WatchDog is able to detect these keywords as soon as 2017, right at their release date but before their disclosure. 
Script \ref{script:Network} shows an example fingerprinting snippet that collects all of the aforementioned network properties.

\textbf{\texttt{Geolocation}} API, standardized in 2008 \cite{geolocation_standardization_2008} and supported in all major browsers around 2009, is a permissioned API that provides information about geographical location of device including (\texttt{latitude}, \texttt{longitude}, \texttt{altitude}, \texttt{speed}), as well as the accuracy of the acquired location data (\texttt{altitudeAccuracy}), and whenever the position of the device changes (\texttt{watchPosition}).
The information exposed by the Geolocation API can be abused for fingerprinting due to its high precision (a double representing the position in decimal degrees).
Note that the Geolocation API was permissioned from the very beginning because of the obvious privacy concerns acknowledged by W3C \cite{geolocation_standardization_2008}.
%
%
\WatchDog detects these keywords as early as 2011, at the earliest formation of the fingerprinting cluster.
Note that the permission status (i.e., whether or not the user has granted permission) itself reveals one bit of information that can be combined with other fingerprinting features.  
\WatchDog detects the abuse of \texttt{PERMISSION\_DENIED} and \texttt{POSITION\_UNAVAILABLE} in 2016.
Script \ref{script:geolocation} shows a fingerprinting snippet that gathers the aforementioned geolocation information, in addition to other fingerprinting information.

\textbf{\texttt{WebGL}} API, standardized in 2010 \cite{khronos_specification_2011} and supported in all major browsers soon afterwards, is a non-permissioned API that can render interactive3D objects in the browser and manipulate them through JavaScript.
WebGL API can be abused for fingerprinting in two main ways. 
First, WebGL can be used to list all WebGL capabilities to build a fingerprint.
For example, scripts can check for WebGL support using \texttt{window.WebGLRenderingContext} and \texttt{getContext(`webgl')} and list capabilities such as \texttt{SHADING\_LANGUAGE\_VERSION} or \texttt{WEBGL\_debug\_renderer\_info}. 
Second, WebGL can be used to render a canvas image (using \texttt{WebGLRenderingContext.canvas}) that is then encoded and hashed (using \texttt{toDataURL}) to build a fingerprint.
The rendering varies across devices due to differences in the rendering pipeline that involves the operating system, web browser, rendering engine, graphics driver, and the underlying hardware.
Note that WebGL 1 \cite{webgl_specification_webgl1} was extended to WebGL 2 \cite{webgl_specification_webgl2} in 2017 to include new capabilities such as pixel buffer objects (\texttt{GetBufferSubData}), Primitive restart (\texttt{draw\_primitive\_restart}), and rasterizer discard (\texttt{RASTERIZER\_DISCARD}).
%
%
\WatchDog detects the keywords associated with WebGL 1 as early as 2011 and WebGL 2 as early as 2017.

\textbf{\texttt{Performance}} API, standardized in 2011 \cite{performance_timeline_2011} and supported in all major browsers around 2012, is a non-permissioned API that covers \texttt{Performance Timeline}, \texttt{Resource Timing}, \texttt{Performance Timeline}, \texttt{Navigation Timing}, \texttt{Resource Timing}, and \texttt{Paint Timing}.
It allows scripts to accurately measure various performance-related metrics during the page load such as DNS using \texttt{domainLookupStart} and \texttt{domainLookupEnd} or HTTP using \texttt{fetchStart}, \texttt{requestStart}, \texttt{responseStart}, and \texttt{responseEnd}.
However, access to high resolution timing information (up to sub-millisecond granularity) can be abused for fingerprinting by precisely timing certain operations that depend on the underlying software/hardware pipeline \cite{HighResolutionAPIPrivacy}.
For example, in \cite{sanchez2018clock} they measured clock difference on a device using combination of \texttt{Performance} and \texttt{Crypto} API.
Specifically, they used \texttt{performance.now} to time the execution of the pseudo-random generator (\texttt{getRandomValues}) to create a browser fingerprint.
\WatchDog detects most of the associated keywords as early as 2016. 
%
%
\texttt{Paint Timing} API is a recent addition to \textit{Performance} API and has been supported by Chrome since 2017 and in other major browsers since 2020 \cite{PerformancePaintTiming}. 
This API measures the time it takes between the moment a user navigate to a URL and the moment a pixel renders on a screen (e.g., \texttt{first-paint} or \texttt{first-contentful-paint} representing time between navigation start \texttt{performanceEntry.startTime} and when the browser renders any/content pixel, respectively).
This timing information can be distinctive across different browsers based on differences in their underlying compute/communication performance. 
\WatchDog captures the abuse of Paint Timing API in 2019, the first time it appears in our data. 
Script \ref{script:performance} shows a fingerprinting snippet that measures the First Time to Paint and First Contentful Paint in addition to other fingerprinting information.

\begin{figure}[!htpb] 
\begin{lstlisting}[basicstyle=\linespread{0.2},style=htmlcssjs,caption={Simplified version of a script that uses the \texttt{Visibility} API to conduct ephemeral fingerprinting. Each time the visibility state changes, it is recorded with the current timestamp.}, label={script:pagevisibility}]
...
// Register an event that will be 
// triggered on visibility state change.
document.addEventListener('visibilitychange', VisibilityStateHandler);

// return visibility state of the page
function getVisibilityState() {
    return document.visibilityState;
}

// return current time 
function getCurrentTime() {
    return Date.now();
}

// Capture current time & visibility state.
function VisibilityStateHandler() { 
    ...
    VisibilityStateFP = {
      VisibilityState: getVisibilityState(),
      CurrentTime: getCurrentTime()
    };
    ...
}
...
    \end{lstlisting}
    \vspace{-10pt}
\end{figure}

\textbf{\texttt{Page Visibility}} API, standardized in 2011 \cite{pagevisibility2011} and supported in all major browsers by 2013, is a non-permissioned API. 
This API provides access to determine the visibility state (i.e. \texttt{visible}, \texttt{hidden}, and \texttt{prerender}) or be notified when the visibility state of a document changes.
While the visibility state (or the change in visibility state) is not directly useful for fingerprinting, but it can be abused for \texttt{ephemeral fingerprinting} \cite{Ephemeralpagevisibility2020} when the changes in page visibility state can be correlated across different sites.
Specifically, when a user switches between a pair of tabs/windows then a change in the visibility state will be simultaneously triggered for both tabs/windows. 
This information can be correlated by a script on both tabs/windows to link whether the tabs/windows are on the same browser/device. 
For example, Script \ref{script:pagevisibility} measures timestamps of the changes in page visibility state. 
It uses \texttt{Date.now} to log the exact time the page visibility state changes (\texttt{onvisibilitychange}).
The sequence of timestamps when the page visibility state changes is expected be the same and distinctive across all of the co-visible sites in a user's browser/device. 
Thus, it can be used to build a cross-site ephemeral fingerprint.
Disclosed in 2020 \cite{Ephemeralpagevisibility2020}, \WatchDog first detects the abuse of this API in 2017.

\textbf{\texttt{Web Worker}} API, standardized in 2009 \cite{WebWorkers2009} and supported in all major browsers by 2010, is a non-permissioned API. 
This API allows sites to run heavy processing scripts in the background without affecting the performance of the main page.
Although DOM and Window objects are not accessible to \texttt{Web Worker} API, however, they do have access to a number of other APIs including \texttt{WebGL}. 
\texttt{Workers} can be used to run a fingerprinting technique (e.g., Canvas fingerprinting using \texttt{OffscreenCanvas} \cite{OffscreenCanvas}) in a background thread separate from the main execution thread of a web application without making the main thread slow or blocked.  
We have not detected such a scenario in our dataset of scripts.
However, \WatchDog detects the presence of this API as early as 2011 where scripts simply probe the support status of this API (e.g., using \texttt{window.Worker}) alongside other fingerprinting information.
\WatchDog also detects \texttt{SharedWorkerGlobalScope.applicationCache} in a number of scripts as the cache of the \texttt{worker} that allow scripts to set and get client-side state as an alternative to cookies.

%

\textbf{\texttt{Gamepad}} API, standardized in 2014 \cite{Gamepad2014} and supported in all major browsers since then is a non-permissioned API that allows browsers to connect to gamepads.
\texttt{getGamepads} method returns the list of \texttt{Gamepad} objects as well as their configuration such as \texttt{axes},  \texttt{buttons}, \texttt{displayId} or \texttt{hand}.
Probing whether a browser has a connected \texttt{Gamepad} and, if there is one connected, collecting the aforementioned configuration information can reveal distinctive information about a browser.
Due to its potential privacy threats, starting 2020, Mozilla requires thirds-party iframes to ask for permission before calling the \texttt{getGamepads} method \cite{securinggamepad2020}.
\WatchDog detects these keywords as early as 2014, right after it was supported in major browsers, even though it was disclosed 6 years later.
%
%
Script \ref{script:gamepad} shows a fingerprinting snippet from 2014 that probes the presence of \texttt{Gamepad} API and calling the \texttt{getGamepads} method in addition to collecting other fingerprinting information.

\textbf{\texttt{Mouse}}-related interfaces, including \texttt{MouseEvent}, \texttt{WheelEvent}, \texttt{MouseScrollEvent}, \texttt{MouseWheelEvent}, and \texttt{Pointer Lock}, was first introduced in 2004 and has since been updated to support new features.
It can capture coordinates of a pointing device's (such as a mouse) including \texttt{clientX/Y}, \texttt{pageX/Y}, \texttt{offsetX/Y}, \texttt{movementX/Y} in addition to its events such as \texttt{click}, \texttt{dblclick}, and \texttt{mousemove} without granting any permission.
Beginning as early as 2004 \cite{pusara2004user}, there has been a steady stream of studies demonstrating how mouse movements can be used to identify users \cite{thomas2021broad}.  
%
%
\WatchDog first detects the abuse of mouse-related keywords for user behavior fingerprinting in 2016 and since then has detected other properties such as \texttt{movementX/Y}, \texttt{deltaX/Y/Z}, and \texttt{wheelDelta}.
Script \ref{script:mouse} shows an example fingerprinting snippet that collects mouse movement information in addition to other fingerprinting information.

\textbf{\texttt{TouchEvent}} interface, standardized in 2011 \cite{TouchEvents2011} and supported by all major browsers including mobile version of browsers since 2013.\footnote{Desktop version of Firefox started supporting this interface lately in 2017 \cite{TouchEvent}.}
\texttt{TouchEvent} is a non-permissioned interface that is similar to mouse interfaces except that it supports simultaneous touches and at different locations on the touch surface.
Beginning as early as 2013 \cite{shahzad2013secure}, there has been a steady stream of studies demonstrating how touch events can be use to identify users \cite{thomas2021broad}.  
Specifically, frequency of tapping (captured by events such as \texttt{ontouchstart}, \texttt{touchenter}, \texttt{touchleave}, and \texttt{touchmove}), and strength of tapping (captured by \texttt{force}) can be used for user behavior fingerprinting.
\WatchDog first detects the abuse of touch-related keywords for user behavior fingerprinting in 2011 and since then has detected other properties such as \texttt{rotationAngle}.

\textbf{\texttt{Sensor}} APIs, standardized in 2012 \cite{SensorAPI2012} and is only supported in Chrome since 2017.
Privacy-oriented browsers like Firefox and Safari have declined to implement this API due to privacy concerns \cite{apple_declined_16_apis, SensorAPIsMDN}.
It is a permissioned API that provides sensor information such as light intensity (using \texttt{AmbientLightSensor}) and the force caused by vibration or a change in motion (using \texttt{Accelerometer}.
Older interfaces such as \texttt{DeviceMotionEvent} and \texttt{DeviceOrientationEvent}, which are not part of \texttt{Sensor} AIP but implemented by all major browsers (except Safari) since 2011 \cite{DeviceMotionEvent}, provide non-permissioned access to a subset of sensors related to a device's position and orientation.
The information exposed by these APIs and interfaces has been shown to be used for user behavior fingerprinting \cite{van2016accelerometer, bojinov2014mobile, thang2012gait}.
\WatchDog detects the sensor keywords associated with \texttt{DeviceMotionEvent} starting from 2017.
Although in the previous studies, the sensor data was collected using \texttt{DeviceMotionEvent}, we detect the abuse of \texttt{Sensor} API that are not supported by \texttt{DeviceMotionEvent} and \texttt{DeviceOrientationEvent}.
For example, \WatchDog detects the abuse of \texttt{AmbientLightSensor} and \texttt{illuminance} that are not yet disclosed. 
Script \ref{script:sensor} shows an example fingerprinting snippet that collects sensor information in addition to other fingerprinting information.
%

\textbf{\texttt{Clipboard}} API, standardized in 2015 \cite{Clipboard2015} but not supported as early as 2018 by major browsers \cite{ClipboardAPI}, implements clipboard operations such as \texttt{copy}, \texttt{cut}, and \texttt{paste}.
Moreover, if a user grants permission, it provides asynchronous access to read and modify the contents of the system clipboard using \texttt{read} (or \texttt{readText} or \texttt{clipboardData.getData('Text')}) and \texttt{write} (or \texttt{writeText}) methods.
However, since this API can access the clipboard data, there are serious privacy concerns due to the possibility that the clipboard might contain personally identifiable information (PII) \cite{AsyncClipboard}.
\WatchDog detects the abuse of Clipboard API as early as 2018. 
Script \ref{script:clipboard} shows an example fingerprinting snippet that collects clipboard information in addition to other fingerprinting information.

\begin{figure}[ht]
  \centering
  \includegraphics[width=\linewidth]{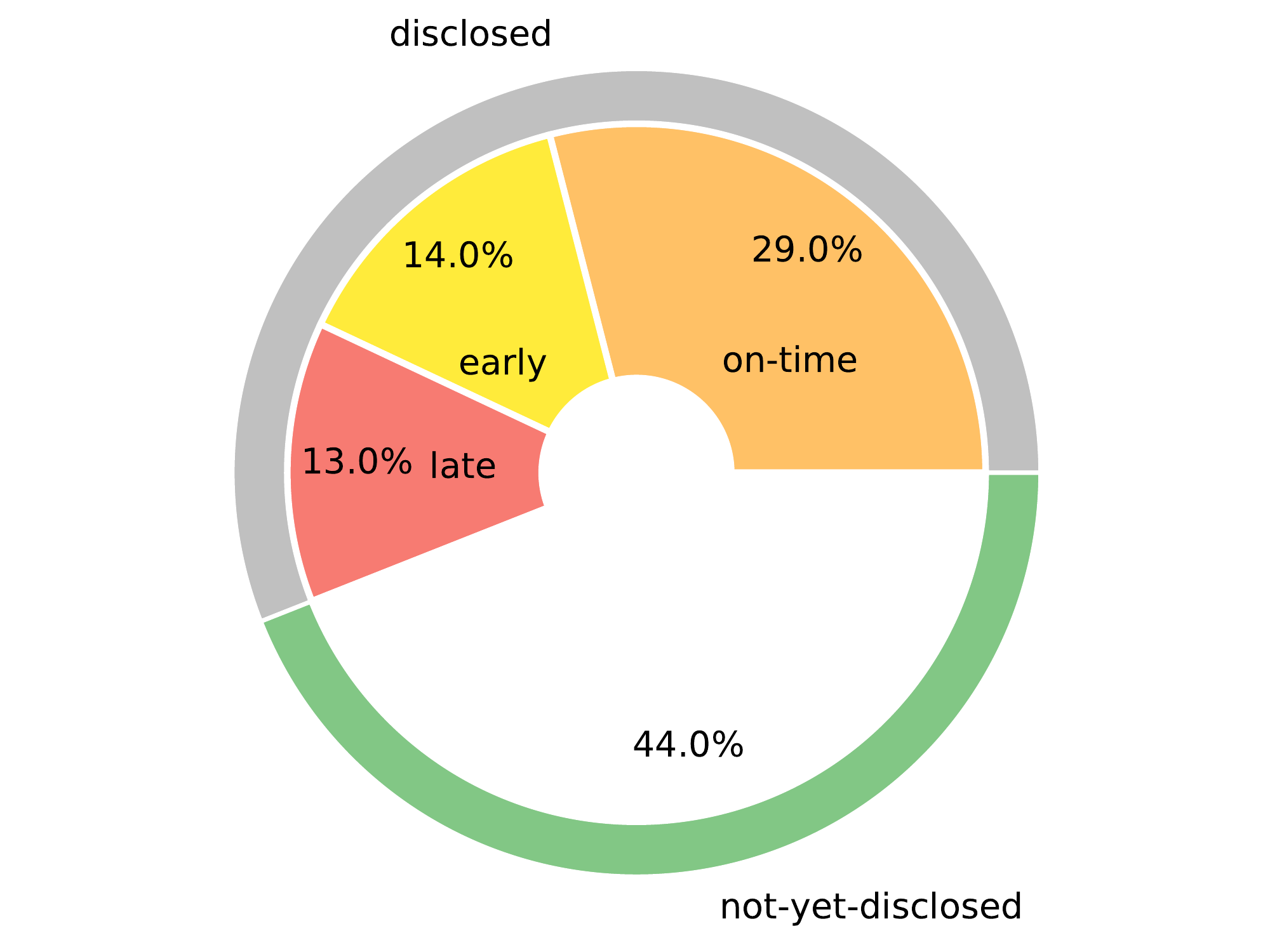}
  \caption{Breakdown of disclosed and undisclosed APIs along with \name's detection time.}
  \label{fig:detection-time}
\end{figure}

\textbf{Validation.}
We sift through public disclosures to validate the fingerprinting potential and abuse of APIs listed in Table \ref{tab: APIs}.
Figure \ref{fig:detection-time} shows the breakdown of disclosed and undisclosed APIs along with \name's detection time. 
%
We find that the 44\% of API keywords detected by \name are still publicly undisclosed. 
%
We try to validate the remaining undisclosed detections by comparing with  DuckduckGo's recently released list of fingerprinting APIs \cite{ddg_fp_discoveries}.  
We note that \name detects more than 80\% of fingerprinting API keywords detected by DuckDuckGo.
However, 90\% of keywords detected by \name, including several well-known fingerprinting APIs, such as \texttt{Battery.changingTime}, \texttt{Geolocation.geolocation}, and \texttt{WebGL.WEBGL\_debug\_renderer\_info} are still undetected by DuckDuckGo. 
DuckDuckGo primarily misses the remaining keywords because it detects a limited number of APIs, i.e., 96, using a very simple heuristic that uses the ratio of an APIs appearance in ``suspicious'' script to an ``non-suspicious'' script to label them as fingerprinting (more details in Section \ref{sec: related work}). 
%

%

%
%

\section{Limitations}
\label{sec: limitations}
In this section, we discuss some of the limitations of \name's pipeline including completeness of retrospective measurements, robustness of the analysis technique, and ground truth assessment of fingerprinting scripts and fingerprinting techniques.

\vspace{.05in} \noindent \textbf{Measurements.}
\name relies on the Wayback Machine for retrospective longitudinal measurements of browser fingerprinting. 
%
%
As we discuss in Section \ref{section:longitudional-data-crawling}, the Wayback Machine's archiving process has limitations that lead to potentially incomplete coverage.
Unfortunately, to the best of our knowledge, there is no other publicly available service that archives complete historical version of webpages. 
HTTP Archive \cite{http_archive} is a related project that archives millions of URLs each month. 
However, it does not store the response bodies of all of the resources \cite{httparchivequickstart} and the downloadable data is only available for the last 6 year, i.e., 2016 to 2021 \cite{http_archive_data}.
Given the democratization of large-scale web crawling tools and capabilities, going forward, future work can consider conducting live crawls to complement missing resources in archiving services such as the Wayback Machine or HTTP Archive.

\vspace{.05in} \noindent \textbf{Robustness.}
\name relies on static analysis of JavaScript code snippets, i.e., AST-based representation of scripts, to extract web API keywords. 
Relying on static analysis makes it challenging for \name to process obfuscated scripts and attribute some generic keywords to APIs. 
Specifically, some fingerprinting scripts use \texttt{eval}-based code obfuscation techniques \cite{Skolka19minifiedobfuscatedJS} and some keywords are implemented by multiple APIs, e.g., \texttt{font} is implemented by \texttt{CanvasRenderingContext2D} \cite{canvasrenderingcontext_font} and \texttt{HTMLElement.style} \cite{htmlelement_style}. 
We attempt to unpack obfuscated scripts by loading them in an instrumented browser and extracting scripts as they are parsed by the JavaScript engine. 
This approach is able to unpack scripts containing \texttt{eval} or \texttt{Function}, but does not fully address other more sophisticated obfuscation techniques \cite{sarker20imcobfuscation}. 
While we do not fully address the keywords attribution issue because it is non-trivial to attribute a generic keyword to the calling API without executing the scripts, we mitigate this issue in our analysis by filtering generic  keywords that belong to multiple APIs.
To fully address these concerns, \name can be extended to include dynamic analysis as well; however, it suffers from code coverage issues that are non-trivial to address. 
Note that \name is not susceptible to obfuscation by a small number of scripts since it leverages tens of thousands of scripts to build its API keyword co-occurrence graph representation. 
Similarly, filtering a small number of generic keywords does not affect the correctness of the analysis.

\vspace{.05in} \noindent \textbf{Ground truth.}
Since \name uses unsupervised clustering, it  relies on the classification of fingerprinting scripts \cite{fpinspector_CodeData}, provided by FP-Inspector \cite{iqbal21fpinspector}, to label the fingerprinting cluster. 
Since the classifications of FP-Inspector are not validated for scripts observed in prior years, we cannot solely rely on that as our ground truth while investigating  fingerprinting techniques in Section \ref{sec: ResultsDiscussion}.
To mitigate this concern, we conduct manual analysis to validate the fingerprinting abuse of the APIs detected by \name. 
We also rely on a wide range of additional external sources including W3C documents, published research papers, and bug reports to assist with our manual analysis.



\section{Conclusion}
We presented \name, a machine learning approach for early detection of web API abuse for browser fingerprinting. 
\name advances the state-of-the-art in browser fingerprinting in two major ways. 
First, unlike prior work that is limited to analyzing a specific fingerprinting technique(s) at a particular point in time, \name conducts a retrospective longitudinal measurement study of browser fingerprinting over the last decade using the Wayback Machine. 
Second, unlike prior work that is limited to detecting deployment of already known fingerprinting techniques, \name is able to detect abuse of new methods of existing web APIs or new web APIs altogether by leveraging the aforementioned longitudinal measurements to model and cluster the evolution of API usage as a temporal graph.
Most notably, \name detects novel types of user environment/hardware fingerprinting such as peripheral configuration via \texttt{Gamepad} and system capabilities via \texttt{Network Information} APIs as well as ephemeral fingerprinting of \texttt{Page Visibility} API even though it does not directly expose highly identifying information.

\name is able to detect the abuse of web API features before/at their disclosure, thus demonstrating its utility as an early detection system that can inform standards bodies and browser vendors interested in designing and deploying mitigations in a timely fashion.  
\name can complement prior approaches (e.g.,  \cite{Englehardt16MillionSiteMeasurementCCS,iqbal21fpinspector,rizzo2021unveiling}) to detect fingerprinting scripts by helping them adapt to new and evolving fingerprinting techniques. 
In addition to disclosing our findings to relevant entities, we plan to release \name's code and longitudinal measurement dataset artifacts to foster follow-up research.
We also plan to collaborate with existing web tracking projects (e.g., \cite{ddg_tracker_radar_fp,princeton_webtap}) to develop a public-facing implementation that can leverage their live web crawls in the future.

\section*{Acknowledgment}
This work is supported in part by the National Science Foundation under grant numbers 2102347, 2051592, 2103439, and 2127309 (Computing Research Association for the CIFellows 2021 Project).

\bibliographystyle{abbrv}
\bibliography{bib}


\section{Appendix}
\label{appendix}

We provide examples of actual fingerprinting snippets to support the discussion in the main text. 
We make minor revisions to the code to improve its readability.
Note that all of the code snippets provided here use multiple fingerprinting techniques.
However, we only show the relevant part of the code that is pertinent to our discussion in the main text.

\tiny

\begin{figure}
\begin{lstlisting}[basicstyle=\linespread{0.2},basicstyle=\tiny,style=htmlcssjs,caption=Simplified version of a script that uses the \texttt{Battery status} API for fingerprinting., label={script:battery_status}]
// Battery Status API support probing.
if ('getBattery' in navigator) {
  BatteryManagerObj=navigator.getBattery() 
  || navigator.battery()
  BatteryManagerObj.then(monitorBattery);
} 
else {
  ChromeSamples.setStatus('not supported');
}

// Get battery level, charging, 
// and discharging time.
function getStatus(battery) {
  return Math.floor(100 * battery.level)
}

// Trigger the function whenever 
// the battery status changes.
function monitorBattery(battery) {
  // get the battery level
  getStatus(battery);

  // Monitor for further updates.
  ["chargingchange","chargingtimechange",
  "dischargingtimechange", "levelchange"].
  forEach(function(battery) {
    a.addEventListener(battery, null)
  })
}
    \end{lstlisting}
\end{figure}

\begin{figure}
\begin{lstlisting}[basicstyle=\linespread{0.2},basicstyle=\tiny,style=htmlcssjs,caption=Simplified version of a script that reads several of the \texttt{Navigator} API properties to conduct fingerprinting., label={script:navigator}]
function getUseragentData(t) {
  nvgtr_dict = {},
  nvgtr_dict.PX59 = navigator.userAgent,
  nvgtr_dict.PX61 = navigator.language,
  nvgtr_dict.PX313 = navigator.languages, 
  nvgtr_dict.PX63 = navigator.platform,
  nvgtr_dict.PX86 = !!(navigator.doNotTrack || null === navigator.doNotTrack || navigator.msDoNotTrack || window.doNotTrack),
  nvgtr_dict.PX88 = getMimeType(),
  nvgtr_dict.PX169 = navigator.mimeTypes && navigator.mimeTypes.length || -1, 
  nvgtr_dict.PX62 = navigator.product,
  nvgtr_dict.PX69 = navigator.productSub, 
  nvgtr_dict.PX64 = navigator.appVersion;
  nvgtr_dict.PX65 = navigator.appName    
  nvgtr_dict.PX66 = navigator.appCodeName    
  nvgtr_dict.PX67 = navigator.buildID
  nvgtr_dict.PX51 = navigator.plugins,
  nvgtr_dict.PX60 = "onLine" in navigator && !0 === navigator.onLine,
  nvgtr_dict.PX68 = "cookieEnabled" in navigator && !0 === navigator.cookieEnabled   }

function getMimeType() {
  try {
    var t = navigator.mimeTypes && navigator.mimeTypes.toString();
    return "[object MimeTypeArray]" === t || /MSMimeTypesCollection/i.test(t) }
  catch (t) { return !1}    }
    \end{lstlisting}
\end{figure}

\begin{figure}
\begin{lstlisting}[basicstyle=\linespread{0.2},style=htmlcssjs,caption=Simplified version of a script that uses several properties of the \texttt{Network Information} API to conduct fingerprinting., label={script:Network}]
function NetworkConnection(i) {
 function connectionObject(t, i, r) {
  // Returns the NetworkInformation object 
  // that contains information about the  
  // network connection of a device.
  return navigator.connection || navigator.mozConnection || navigator.webkitConnection
 }
 
 // Return network properties.
 return t(a, i.Events), r(a, [{
  value: function () {
  this._dataQueue.addToQueue(
  ...
  timestamp: this.getEventTimestamp(),
  connectionType: this._connection.type ? 
  this._connection.type : "",
  efectivType:this._connection.effectiveType
  ? this._connection.effectiveType : "",
  downlinkMax: this._connection.downlinkMax
  ? this._connection.downlinkMax.toString()
  : "",downlink: this._connection.downlink ?
  this._connection.downlink.toString() : "",
  rtt: this._connection.rtt ?
  this._connection.rtt.toString() : "",
  ...
  )}
  ...
}
    \end{lstlisting}
\end{figure}

\begin{figure}
\begin{lstlisting}[basicstyle=\linespread{0.2},style=htmlcssjs,caption={Simplified version of a script that collects email hashes, does cookie matching and uses \texttt{Geolocation} API to conduct fingerprinting.}, label={script:geolocation}]
// Other tracking functionality. 
...
this.monitorEmailHashes = function() {...},
this.doCookieMatching = function() {...},


// Collection of latitude and longitude.
this.requestGeo = function() {
var e = this;
navigator.geolocation.getCurrentPosition(
  function(t) {
  e.bountyAppend("lat",t.coords.latitude),
  e.bountyAppend("lng",t.coords.longitude),
  e.bountyAppend("acc",t.coords.accuracy)
}, function(t) {
  e.error("Could not lookup Geo Location")
}, {
    enableHighAccuracy: !0,
    timeout: 1500,
    maximumAge: Infinity
})
},

// Collection of other fingerprinting information. 
this.collectBrowserInfo = function() {...}
...
    \end{lstlisting}
\end{figure}

\begin{figure}
\begin{lstlisting}[basicstyle=\linespread{0.2},style=htmlcssjs,caption=Simplified version of script that uses the \texttt{Clipboard} API to conduct ephemeral fingerprinting., label={script:clipboard}]
...
// Capturing clipboard text & current time.
u._sendToQueue = function(e, t) {
  var n = u.getEventTimestamp(e),
  o = e.clipboardData ? 
  e.clipboardData.getData("text") : 
  window.clipboardData ? 
  window.clipboardData.getData("text"):"";
   
  var s = u.getExports().
  EnumDef.Events.clipboardEventType[e.type];
  
  u._dataQueue.addToQueue("clipboard_event",  
  {timestamp: n, copiedText: o, 
  clipboardEventType: s})               
  ...
 } 
 ...
    \end{lstlisting}
\end{figure}

\begin{figure}
\begin{lstlisting}[basicstyle=\linespread{0.2},style=htmlcssjs,caption={Simplified version of script that uses the \texttt{Performance} API to conduct fingerprinting.}, label={script:performance}]
...
// Reading the timing of webpage paint events.
{
key: "onWindowLoad",
value: function() {
  y.a.preloadAll();
  e = performance.getEntriesByType("paint");
  
  e.forEach(function(e) {
    console.log("".concat(e.name, ": ").concat(e.startTime))
  })
}
...
    \end{lstlisting}
\end{figure}

\begin{figure}
\begin{lstlisting}[basicstyle=\linespread{0.2},style=htmlcssjs,caption=Simplified version of a script that probes the \texttt{GamePad} API to conduct fingerprinting., label={script:gamepad}]
...
onLoad = function() {
frame = document.createElement('iframe');
flags = [];
...
if (isPresent(navigator, 'getGamepads')) {
    flags.push('gamepads');
}
...
flags = flags.join(',');
frame.src = ("http://" + host + "/statframe.html#") + flags;
frame.style.cssText = 'display: none;';
return document.body.appendChild(frame);
...
}; 
    \end{lstlisting}
\end{figure}

\begin{figure}
\begin{lstlisting}[basicstyle=\linespread{0.2},style=htmlcssjs,caption=Simplified version of a script that uses the \texttt{Mouse} API to conduct fingerprinting., label={script:mouse}]
...
function mn(t) {
g("PX847");
var n = p();
if (va) {
  var e = pa[si];
  ua = si, la = n;
  var r = t.deltaY || t.wheelDelta 
  || t.detail;
  
  if (r = +r.toFixed(2), null === e) {
    fa++;
    var o = wn(t, !1);
    o.PX172 = [r], o.PX173 = gt(n)
    , pa[si] = o
  } 
else ma.mousewheel <= pa[si].PX172.length
? (Xn(), va = !1) : pa[si].PX172.push(r)}

X("PX847")
}

function gn() {
 if (g("PX847"), pa.mousemove) {
  t = pa.mousemove.coordination_start.length
  , n = pa.mousemove.
  coordination_start[t-1].PX70,
  e = Sn(Tn(_t(pa.mousemove.
  coordination_start))),
  r = Tn(_t(pa.mousemove.coordination_end));
  r.length > 0 && (r[0].PX70 -= n);
  var o = Sn(r);
  
   pa.mousemove.PX172 = "" !== o 
   ? e + "|" + o : e, 
   delete pa.mousemove.coordination_start, 
   delete pa.mousemove.coordination_end, 
   yn(pa.mousemove, "mousemove"),
   pa.mousemove = null
 }
 X("PX847")
}
...
    \end{lstlisting}
\end{figure}

\begin{figure}
\begin{lstlisting}[basicstyle=\linespread{0.2},style=htmlcssjs,caption=Simplified version of a script that uses \texttt{Sensor} APIs to conduct fingerprinting., label={script:sensor}]
...
vn.DataMappingDefs = {  
// This script define 23 variable with the name of methods/properties related to each API. Then starts collecting information for each API including Sensor API.
...
AMBIENT_LIGHT_EVENT_MAP: ["eventSequence", "timestamp", "illuminance"],
ACCELEROMETER_EVENT_MAP: ["eventSequence", "timestamp", "x", "y", "z"],
GYRO_EVENT_MAP: ["eventSequence", "timestamp", "absolute", "alpha", "beta", "gamma"],
ORIENTATION_EVENT_MAP: ["eventSequence", "timestamp", "absolute", "alpha", "beta", "gamma"],
...
    \end{lstlisting}
\end{figure}

\end{document}